\documentclass[twocolumn,tighten]{aastex631}
\usepackage{amsfonts,amsmath,amssymb}
\usepackage{bm}
\usepackage{xcolor}
\usepackage{subfigure}
\usepackage[]{cleveref}
\usepackage{mathtools,upgreek,pifont}
\usepackage{cuted}

\newcommand{\imag}{{\rm i}}
\newcommand{\rmd}{{\rm d}}
\newcommand{\rme}{{\rm e}}
\newcommand{\pD}[2]{\frac{\partial #2}{\partial #1}}

\newcommand{\D}[2]{\frac{\rmd #2}{\rmd #1}}
\newcommand{\DD}[2]{\frac{\rmd^2 #2}{\rmd {#1}^2}}

\newcommand\bb[1]{\mbox{\boldmath{$#1$}}}
\newcommand\grad{\bb{\nabla}}
\newcommand\bcdot{\,\bb{\cdot}\,}

\newcommand\btimes{\,\bb{\times}\,}

\newcommand{\ez}{\hat{\bb{z}}}

\newcommand{\ey}{\hat{\bb{y}}}

\shorttitle{Tearing Onset in Poorly Ionized Plasmas}
\shortauthors{Tolman et al.}

\begin{document}
\title{Tearing-mediated reconnection in magnetohydrodynamic poorly ionized plasmas. \\ I. Onset and linear evolution}

\author[0000-0002-2642-064X]{Elizabeth A.~Tolman}
\affiliation{Institute for Advanced Study, 1 Einstein Drive, Princeton, NJ, 08540, USA}
\affiliation{Center for Computational Astrophysics, Flatiron Institute, 162 Fifth Avenue, New York, NY, 10010, USA}
\correspondingauthor{Elizabeth A.~Tolman}
\email{etolman@flatironinstitute.org}

\author[0000-0001-7801-0362]{Matthew W.~Kunz}
\affiliation{Department of Astrophysical Sciences, Princeton University, 4 Ivy Lane, Princeton, NJ 08544, USA}

\author[0000-0002-0067-1272]{James M.~Stone}
\affiliation{Institute for Advanced Study, 1 Einstein Drive, Princeton, NJ, 08540, USA}

\author[0000-0002-0067-1272]{Lev Arzamasskiy}
\affiliation{Institute for Advanced Study, 1 Einstein Drive, Princeton, NJ, 08540, USA}

\begin{abstract}
In high-Lundquist-number plasmas, reconnection proceeds via onset of tearing, followed by a nonlinear phase during which plasmoids continuously form, merge, and are ejected from the current sheet (CS). This process is understood in fully ionized, magnetohydrodynamic plasmas. However, many plasma environments, such as star-forming molecular clouds and the solar chromosphere, are poorly ionized. We use theory and computation to study tearing-mediated reconnection in such poorly ionized systems.  In this paper, we focus on the onset and linear evolution of this process. In poorly ionized plasmas, magnetic nulls on scales below $v_{\rm A,n0}/\nu_{\rm ni0}$, with $v_{\rm A,n0}$ the neutral Alfv\'{e}n speed and $\nu_{\rm ni0}$ the neutral--ion collision frequency, will self-sharpen via ambipolar diffusion.  This sharpening occurs at an increasing rate, inhibiting the onset of reconnection. Once the CS becomes thin enough, however, ions decouple from neutrals and thinning of the CS slows, allowing tearing to onset in a time of order $\nu_{\rm ni0}^{-1}$.  We find that the wavelength and growth rate of the mode that first disrupts the forming sheet can be predicted from a poorly ionized tearing dispersion relation; as the plasma recombination rate increases and ionization fraction decreases, the growth rate becomes an increasing multiple of $\nu_{\rm ni0}$ and the wavelength becomes a decreasing fraction of $v_{\rm A,n0}/\nu_{\rm ni0}$.

\end{abstract}

\keywords{}

\section{Introduction}
Magnetic reconnection \citep{zweibel2009magnetic,loureiro2015magnetic} is a fundamental plasma process that can occur in a broad set of terrestrial and astrophysical plasmas \citep{masuda1994loop, hender2007mhd,mallet2017disruption,loureiro2017role,boldyrev2017magnetohydrodynamic,dong2018role,walker2018influence,loureiro2020nonlinear,chernoglazov2021dynamic,galishnikova2022tearing,philippov2019pulsar,ripperda2020magnetic}. In plasmas of high Lundquist number $S$, where $S\equiv L_{\rm CS} v_{\rm A0}/\eta $ with $L_{\rm CS}$ the length of the current sheet (CS), $v_{\rm A0}$ the Alfv\'{e}n speed, and $\eta$ the plasma resistivity, reconnection proceeds through the onset of the tearing instability, followed by a complicated nonlinear phase in which plasmoids form, merge, and are advected along a CS   \citep{loureiro2007instability,uzdensky2010fast,ji2011phase}. 

Theoretical understanding of reconnection in high-Lundquist-number, fully ionized, magnetohydrodynamic (MHD) plasmas has recently approached maturity. Pioneering works have considered the linear growth of the tearing instability in static CSs \citep{furth1963finite,coppi1976resistive} and in dynamically forming CSs \citep{pucci2013reconnection,uzdensky2016magnetic,comisso2016general,huang2017plasmoid,tolman2018development}\footnote{Many plasma environments are not sufficiently collisional to be described by MHD and require kinetic treatments. Recent works in this area involving dynamically forming CSs and tearing onset include \cite{alt2019onset} and \cite{winarto2021triggering}.} and the resulting nonlinear plasmoid chain \citep{shibata2001plasmoid,samtaney2009formation,uzdensky2010fast,loureiro2012magnetic}.  Evidence supporting an emerging theoretical picture in which reconnection in a forming CS is initiated by strongly unstable tearing modes that lead to a nonlinear plasmoid chain has been found in astrophysical observations \citep{nishizuka2010multiple,takasao2011simultaneous} as well as in targeted laboratory experiments \citep{hare2017anomalous}.

However, many plasmas of current interest in the astrophysics and laboratory plasma physics communities are not fully ionized, and reconnection may proceed in a different manner in these plasmas.  For example, star-forming molecular clouds have ionization fractions ranging from ${\sim}10^{-9}$ to ${\sim}10^{-4}$  \citep{caselli1998,goicoechea2009ionization}, the diffuse interstellar medium has an ionization fraction that ranges from ${\sim}10^{-4}$ to ${\sim}1$ \citep{heiles2012magnetic,fielding2022plasmoid}, protoplanetary disks have ionization fractions that vary widely about ${\sim}10^{-11}$ \citep{oberg2011ionization,Lesur2021},  the solar chromosphere plasma has an ionization fraction ${\sim}10^{-4}$--1 \citep{leenaarts2007non}, and certain laboratory reconnection experiment plasmas have ionization fractions as low as almost $10^{-2}$ \citep{lawrence2013laboratory}. Seminal theoretical work on reconnection in poorly ionized environments is several decades old \citep{zweibel1989magnetic,brandenburg1994formation,brandenburg1995effects}. More recent theoretical studies \citep{heitsch2003fast,leake2013magnetic,singh2019effect,ni2020magnetic,murtas2021coalescence}  of poorly ionized reconnection have been less plentiful and have usually not made contact with the conceptual advances developed in recent fully-ionized studies (\cite{pucci2020tearing} does make this contact).  

Thus, in this and an accompanying paper \citep{tolman23II}, we present analytic theory, verified by high-resolution simulations, of tearing-mediated reconnection in a magnetohydrodynamic, poorly ionized plasma; this theory is of the style used in recent fully-ionized studies.  In this paper we study the plasma motion that triggers reconnection, the onset of tearing, and its linear phase of evolution.  Future work \citep{tolman23II} will investigate the nonlinear phase of plasmoid-mediated reconnection.
We consider a scenario for triggering of reconnection that is expected to describe approximately the onset of reconnection in several poorly ionized plasmas. In particular, we assume that some bulk motion in the neutral gas creates a region where the magnetic field reverses over a sufficiently short lengthscale. The associated CS will rapidly self-sharpen further via ambipolar diffusion, a diffusive process existing in poorly ionized plasmas due to the frictional drag between the ionized and the neutral species. This self-sharpening ultimately triggers reconnection. 

The physics of the self-sharpening process determines how reconnection onsets. Models assumed for fully ionized CS formation often have a constant formation rate~\citep[e.g.,][]{uzdensky2016magnetic}. In contrast, ambipolar sharpening of the CS causes the formation rate to accelerate, such that the tendency to tear is never able to overwhelm ambipolar sharpening. As a result, tearing cannot onset during the main CS formation process. However, at sufficiently narrow CS widths, ions completely decouple from the neutrals and the thinning of the CS slows~\citep{brandenburg1995effects}, allowing tearing onset. This onset occurs after a time of order the inverse neutral--ion collision frequency, $\nu_{\rm ni0}^{-1}$.  Analysis of this onset stage shows that the wavelength and growth rate of the mode that first disrupts the forming sheet can be predicted from a poorly ionized tearing dispersion relation; as the plasma recombination rate increases and ionization fraction decreases, the growth rate becomes an increasing multiple of $\nu_{ni0}$ and wavelength becomes a decreasing fraction of $v_{\rm A,n0}/\nu_{\rm ni0}$, where $v_{\rm A,n0}$ is the neutral Alfv\'{e}n velocity. After the linear onset, the system ultimately settles into a nonlinear phase characterized by a stochastic plasmoid chain. This phase is analyzed in~\cite{tolman23II}.

Our analysis begins in Section~\ref{sec:sys} with a specific statement of the plasma physics problem we consider and the analytic tools used to study this problem, some of which are discussed more fully in Appendix~\ref{sec:stationary}. Section~\ref{sec:natform} considers how a CS would form in a poorly ionized plasma in the absence of tearing and what steady state it would approach. Certain characteristics of this steady state are derived in Appendix~\ref{sec:deltaprime}. Then, Section~\ref{sec:onset} uses the results of these prior sections to understand how tearing onsets in a poorly ionized, dynamically thinning CS.  Section~\ref{sec:simulation} verifies this analysis using high-resolution numerical simulations. Section~\ref{sec:conc} summarizes the paper, discusses its implications for plasma physics and for astrophysics, and comments on avenues for future work. An additional appendix (\S\ref{sec:typval}) provides useful numbers for placing our results in an astrophysical context.

\section{Formulation of the Problem}
\label{sec:sys}
In this Section, we state the problem we solve in this paper, provide the equations we use to study this problem, describe a common approximation applied to these equations, and present some results from the stationary linear theory of the tearing instability that will be useful later in the paper.

\subsection{Problem Statement}
\label{sec:pstate}
We study the tearing-mediated onset of magnetic reconnection in a plasma comprised of an ionized, quasineutral fluid of low density and a neutral fluid of much higher density. These fluids interact via frictional drag, ionization, and recombination.  The bulk motions and magnetic field configurations that might exist in such a plasma, and the ways in which they might lead to reconnection, vary, but we show in Section~\ref{sec:natform} that the essential elements of the final stages of reconnection onset in any system must be similar.  This implies that a general model for the reconnecting CS should have broad application to a wide variety of poorly ionized systems.

In particular, we consider a CS that forms in a poorly ionized plasma when motions in the bulk neutral fluid move the magnetic field in a way that creates a null about which the magnetic field reverses direction. To study this system, we orient the $y$ axis along the initial magnetic field and place the origin at the null (we do not consider a guide field). That is, 
\begin{equation}
\label{eq:maginit}
    \bb{B} = B_y(x,t)\ey ,~{\rm with}~B_y(0,t)=0.
\end{equation}
This field reverses direction over a length scale $a(t)$. We are interested in how Equation~\eqref{eq:maginit} evolves to become 
\begin{equation}
\label{eq:totb}
    \bb{B}_{\rm tot} = B_y(x,t) \bb{\hat{y}} + \delta \bb{B}(x,y,t),
\end{equation}
where $\delta \bb{B}(x,y,t)$ is a magnetic perturbation driven by the tearing instability.
Throughout this process, the plasma far from the CS is taken to have constant magnetic field $B_0$ and constant ionized and neutral fluid densities, $\rho_{\rm i0}$ and $\rho_{\rm n0}$ respectively, with their ratio determined by ionization equilibrium.

\subsection{Two-Fluid MHD Equations}
\label{sec:2flu}

We model the poorly ionized plasma that hosts the reversing magnetic field using a system of hydrodynamic and MHD equations that are coupled together by frictional drag from ion--neutral collisions, by ionization of the form that would be caused by cosmic rays, and by recombination. A discussion of this type of model can be found in \cite{draine1986multicomponent}.

The neutral fluid is governed by a momentum equation,
\begin{align}\label{eq:momn}
    \pD{t}{(\rho_{\rm n}\bb{v}_{\rm n})} &+ \grad\bcdot(\rho_{\rm n}\bb{v}_{\rm n}\bb{v}_{\rm n}) = -\grad P_{\rm n} \nonumber\\*
    \mbox{} &- \nu_{\rm ni}\rho_{\rm n}(\bb{v}_{\rm n}-\bb{v}_{\rm i}) - \xi\rho_{\rm n}\bb{v}_{\rm n} + \alpha \rho^2_{\rm i}\bb{v}_{\rm i},
\end{align}
and a continuity equation,
\begin{equation}\label{eq:contn}
    \pD{t}{\rho_{\rm n}} + \grad\bcdot(\rho_{\rm n}\bb{v}_{\rm n}) = -\xi\rho_{\rm n} + \alpha\rho^2_{\rm i} ,
\end{equation}
where $\rho_{\rm n}$ is the neutral fluid mass density, $\bb{v}_{\rm n}$ is the neutral fluid velocity, $P_{\rm n}$ is the neutral fluid pressure, $\rho_{\rm i}$ is the ionized fluid mass density, $\bb{v}_{\rm i}$ is the ionized fluid velocity, and $\nu_{\rm ni}$ is the collision frequency describing the rate of collisional momentum transfer from the neutral fluid to the ionized fluid. The latter results primarily from collisions between the ions and the neutrals, with
\begin{equation}\label{eq:nuni}
    \nu_{\rm ni} = \frac{\rho_{\rm i}}{\rho_{\rm n}} \nu_{\rm in} = \frac{\rho_{\rm i}\langle\sigma w\rangle_{\rm in}}{m_{\rm i}+m_{\rm n}} ,
\end{equation}
where $m_{\rm i}$ is the mass of the ion species, $m_{\rm n}$ is mass of the neutral species, and $\langle\sigma w\rangle_{\rm in}$ is the mean collisional rate (collisional momentum transfer from the neutral fluid to the electrons occurs at a rate $\nu_{\rm ne}$ that is ${\sim}m_{\rm e}/m_{\rm i}$ slower than $\nu_{\rm ni}$). We take the neutrals to be ionized at a constant rate $\xi$ (e.g., by cosmic rays or photoionization) and produced via recombination (dissociative and/or radiative) with constant coefficient $\alpha$. As the neutral and ionized populations exchange mass through ionization and recombination, their momenta are exchanged as well, a process captured by the final two terms on the right-hand side of Equation~\eqref{eq:momn}.

The momentum and continuity equations for the ionized fluid reflect Newton's third law and total mass conservation, with the former making use of quasi-neutrality between the ions and electrons and including the Lorentz force associated with gradients in the magnetic field $\bb{B}$. In units where the vacuum permeability $\mu_0=1$, these equations are 
\begin{align}\label{eq:momi}
    \pD{t}{(\rho_{\rm i}\bb{v}_{\rm i})} &+ \grad\bcdot(\rho_{\rm i}\bb{v}_{\rm i}\bb{v}_{\rm i}) = -\grad P_{\rm i} + (\grad\btimes\bb{B})\btimes\bb{B} \nonumber\\*
    \mbox{} &-\nu_{\rm ni}\rho_{\rm n}(\bb{v}_{\rm i}-\bb{v}_{\rm n}) + \xi\rho_{\rm n}\bb{v}_{\rm n} - \alpha \rho^2_{\rm i}\bb{v}_{\rm i},
\end{align}
\begin{equation}\label{eq:conti}
    \pD{t}{\rho_{\rm i}} + \grad\bcdot(\rho_{\rm i}\bb{v}_{\rm i}) = \xi\rho_{\rm n} - \alpha\rho^2_{\rm i} .
\end{equation}
The magnetic field evolves according to Faraday's law,
\begin{equation}\label{eq:Faraday}
    \pD{t}{\bb{B}} = -\grad\btimes\bb{E} ,
\end{equation}
with an electric field $\bb{E}$ that accounts for flux freezing in the ionized fluid but for the diffusive effects of a constant Ohmic resistivity $\eta$:
\begin{equation}\label{eq:Ohm}
    \bb{E} = -\bb{v}_{\rm i}\btimes\bb{B} + \eta\grad\btimes\bb{B} .
\end{equation}
This version of Ohm's law neglects Hall drifts between the electron and ion fluids \citep[cf.,][]{malyshkin2011}.

Together, Equations~\eqref{eq:Faraday} and \eqref{eq:Ohm} provide the non-ideal MHD induction equation
\begin{align}\label{eq:induction}
    \pD{t}{\bb{B}} &- \grad\btimes(\bb{v}_{\rm n}\btimes\bb{B}) 
    \nonumber\\*
    \mbox{} &= \grad\btimes\bigl[(\bb{v}_{\rm i}-\bb{v}_{\rm n})\btimes\bb{B}\bigr] + \eta\nabla^2\bb{B}.
\end{align}
In the ideal-MHD limit, in which the fluid is perfectly conducting ($\eta=0$) and the ions and neutrals are collisionally well coupled with negligible relative drift velocity ($\bb{v}_{\rm i}\approx\bb{v}_{\rm n}$), the right-hand side of Equation~\eqref{eq:induction} may be dropped and the magnetic flux is effectively frozen into the bulk neutral fluid. 

We close Equations~\eqref{eq:momn} and \eqref{eq:momi} by adopting isothermal equations of state for both the neutral and ionized fluids,
\begin{equation}
    P_{\rm n} = \rho_{\rm n}C^2_{\rm n} \qquad{\rm and}\qquad P_{\rm i} = \rho_{\rm i}C^2_{\rm i} ,
\end{equation}
with the isothermal sound speeds $C_{\rm n}$ and $C_{\rm i}$ taken to be comparable to one another.

From Equation~\eqref{eq:contn}, we find that the ionized and neutral fluid densities far from the CS (where ionization equilibrium holds) satisfy
\begin{equation}\label{eq:ionequil}
    \frac{\rho_{\rm i0}}{\rho_{\rm n0}} = \sqrt{\frac{\xi}{\alpha\rho_{\rm n0}}} \ll 1.
\end{equation}
The Alfv\'en speed in this predominantly neutral fluid is given by
\begin{equation}\label{eq:vA0}
    v_{\rm A,n0} = \frac{B_0}{\sqrt{\rho_{\rm n0}}} ;
\end{equation}
likewise, the Alfv\'en speed defined using the ionized fluid density is denoted
\begin{equation}
   v_{\rm A,i0} = \frac{B_0}{\sqrt{\rho_{\rm i0}}}. 
\end{equation}
These Alfv\'en speeds may be compared to their corresponding isothermal sound speeds to form the plasma beta parameters
\begin{equation}
    \beta_{\rm n0}  = \frac{2C^2_{\rm n}}{v^2_{\rm A,n0}} \qquad{\rm and}\qquad \beta_{\rm i0}  = \frac{2C^2_{\rm i}}{v^2_{\rm A,i0}}.
\end{equation}
Note that $\beta_{\rm i0}\ll\beta_{\rm n0}$. The neutral--ion collision frequency in the region far from the CS where the ionized fluid density is $\rho_{\rm i0}$ is $\nu_{\rm ni0}$. A particularly important length scale for assessing the importance of ambipolar diffusion is the distance traveled by an (undamped) Alfv\'en wave in the predominantly neutral fluid during one neutral--ion collision timescale, 
\begin{equation}
\label{eq:a0}
    a_0 \equiv \frac{v_{\rm A,n0}}{\nu_{\rm ni0}}.
\end{equation}
The importance of resistivity is quantified using the (time-dependent) Lundquist number defined with the CS width,
\begin{equation}
\label{eq:etanorm}
    S_{a}(t) \equiv \frac{a(t) v_{\rm A,n0}}{\eta},
\end{equation}
with its value when $a(t)=a_0$ being denoted by  $S_{a_0}$ (note that our definition of $S_a$ uses the Alfv\'en speed in the neutrals). All variables unadorned with a subscript ``0'', such as $\nu_{\rm ni}$, $v_{\rm A,n}$, and $v_{\rm A,i}$, refer to those quantities evaluated at a particular location and time.

Typical values for some of the parameters presented in this Section can be found in Appendix~\ref{sec:typval}.

\subsection{Strong-Coupling Approximation}\label{sec:strongcoupling}
At certain moments in our study of this CS, we adopt a simplification commonly used in many MHD treatments of poorly ionized plasmas. Namely, the ion--neutral drift velocity is related directly to the Lorentz force via
\begin{equation}\label{eq:strongcoupling}
    \bb{v}_{\rm i}-\bb{v}_{\rm n} \approx \frac{(\grad\btimes\bb{B})\btimes\bb{B}}{\rho_{\rm n}\nu_{\rm ni}}.
\end{equation}
This simplification, called a ``strong-coupling approximation,'' assumes that a terminal velocity is quickly established as the Lorentz force pushes the ionized fluid through the collisional neutrals. From Equation~\eqref{eq:momi}, it is clear that this relation neglects not only the inertia of the ionized fluid, but also its pressure and the interspecies exchange of momentum due to ionization and recombination. 

The ambipolar magnetic diffusivity implied by Equations~\eqref{eq:induction} and \eqref{eq:strongcoupling} is nonlinear in the magnetic-field strength:
\begin{align}\label{eq:Bstrong}
    \pD{t}{\bb{B}} - \grad\btimes(\bb{v}_{\rm n}\btimes\bb{B}) &-  \eta\nabla^2\bb{B} \nonumber\\*
    \mbox{} &\approx - \grad\btimes \biggl( \frac{B^2}{\rho_{\rm n}\nu_{\rm ni}} \grad\btimes\bb{B} \biggr)_{\!\perp} ,
\end{align}
where the subscript ``$\perp$'' denotes the component perpendicular to the local magnetic field. Indeed, for the CS profile described by Equation~\eqref{eq:maginit}, the right-hand side of Equation~\eqref{eq:Bstrong} becomes
\begin{equation}
\label{eq:bev}
    +\D{x}{} \biggl( \frac{B_y^2}{\rho_{\rm n}\nu_{\rm ni}} \D{x}{B_y} \biggr) .
\end{equation}
In this case, the ion--neutral drift velocity is directly proportional to $-{\rm d}B_y^2/{\rm d}x$, so that the magnetic flux is pushed towards the magnetic null at a rate that increases with $B_y^2$. At the same time, the ambipolar diffusion coefficient $B_y^2/\rho_{\rm n}\nu_{\rm ni}$ vanishes at the center of the sheet, and so the field close to the null is unable to smooth itself out as it steepens. As a result, without finite ionized fluid pressure and Ohmic dissipation, the magnetic gradients sharpen into singularities \citep{brandenburg1994formation}.
In particular, the steady-state solution of Equation~\eqref{eq:Bstrong} is 
\begin{equation}
\label{eq:generalform}
    B_y(x) = C_1 \left( 3x-C_2\right)^{1/3} ,
\end{equation}
with $C_1$ and $C_2$ being arbitrary constants. For the specific situation described by Equation~\eqref{eq:maginit}, Equation~\eqref{eq:generalform} becomes
\begin{equation}
\label{eq:generalform2}
    B_y(x) = C_1 \, 3^{1/3} x^{1/3}.
\end{equation}
The current $\bb{J}$ to which this magnetic field corresponds is given by
\begin{equation}
    \bb{J} =  C_1 \, 3^{-2/3}x^{-2/3} \ez.
\end{equation}
The singularity of this equation at $x=0$ indicates the failure of the strong-coupling approximation; in Section~\ref{sec:pisp} we show how this singularity is resolved by finite ionized fluid pressure and Ohmic resistivity.

\subsection{Tearing Growth Rates in Poorly Ionized Plasma}\label{sec:tearing}
To predict when a forming CS might become disrupted by the tearing instability, we first require a theory for this instability's linear growth rate and characteristic scale in a stationary poorly ionized plasma. That is, we require a theory for $\delta\bb{B}(x,y,t)$ in Equation~\eqref{eq:totb} for the case of $B_y(x,t) = B_y (x)$. This theory will serve in Section~\ref{sec:onset} as a tool to understand the stability of the time-dependent, forming sheet. 

We develop this theory in Appendix~\ref{sec:stationary}; key results are provided here. The perturbed quantities can all be Fourier analyzed, so that the magnetic-field fluctuation is decomposed as
\begin{equation}
\label{eq:eigdef}
    \delta \bb{B} (x,y,t) = \delta \bb{B}(x)\exp(\gamma t +\imag k y),
\end{equation}
where $\gamma$ is the tearing-mode growth rate and $k$ is a wavenumber along the CS. The growth rate pertaining to a fully ionized CS is determined implicitly by the dispersion relation \citep{ coppi1976resistive}
\begin{equation}
\label{eq:gamma}
    \Delta' a = -\frac{\pi}{8} \, \Lambda^{5/6}  \, 
    \dfrac{\Gamma\bigl[(\Lambda-1)/4\bigr]}{\Gamma\bigl[(\Lambda+5)/4\bigr]} \biggl(\frac{\tau_\eta}{\tau_{\rm A,i}}\biggr)^{1/3},
\end{equation}
where $a$ is the characteristic width of the CS. The parameter ${\Delta}'$ is the tearing-mode stability index \citep{furth1963finite}, which is a function of $k$ and $a$ that depends on the functional form of the CS's equilibrium magnetic field. In addition,
\begin{equation}
    \tau_{{\eta}} \equiv a^2/\eta
\end{equation}
is the characteristic timescale on which the CS diffuses resistively, and
\begin{equation}
\label{eq:tauA}
    {\tau}_{\rm A,i} \equiv (kv_{\rm A,i})^{-1}
\end{equation}
is the characteristic timescale for an (undamped) Alfv\'en wave propagating in the ionized fluid at speed $v_{\rm A,i}$ to cross a fluctuation having wavenumber $k$. 
The eigenvalue
\begin{equation}
\label{eq:reslayer}
    \Lambda( \gamma) \equiv \gamma^{3/2} \tau_{\rm A,i} \tau^{1/2}_\eta \equiv \gamma \delta^2_\eta/\eta
\end{equation}
is the growth rate of the tearing mode normalized by the rate of  diffusion across the inner resistive boundary layer of width $\delta_\eta$.

To apply these expressions for a poorly ionized system, we introduce an effective Alfv\'{e}n velocity,
\begin{equation}
\label{eq:sub}
    v'_{\rm A,i} \equiv v_{\rm A,i} \biggl( 1 + \frac{\rho_{\rm n}}{\rho_{\rm i}}\dfrac{\nu_{\rm ni}+\alpha \rho_{\rm i}^2/\rho_{\rm n}}{\gamma+\nu_{\rm ni}+\xi}\biggr)^{-1/2};
\end{equation}
note that this quantity depends upon $\gamma$. It turns out that Equation~\eqref{eq:gamma} applies equally well to the poorly ionized case if we simply replace 
\begin{equation}\label{eq:tauApart}
    {\tau}_{\rm A,i} \,\rightarrow\, {\tau}'_{\rm A,i} \equiv (kv'_{\rm A,i})^{-1} .
\end{equation}
This fact was first pointed out by \citet{zweibel1989magnetic} and has also been used by \citet{pucci2020tearing}. Together, Equations~\eqref{eq:gamma}, \eqref{eq:sub}, and \eqref{eq:tauApart} determine the linear growth rate of the tearing mode in a poorly ionized plasma. 

It will be helpful to have not only the full dispersion relation for the tearing growth rate (Equation~\eqref{eq:gamma} with Equation\eqref{eq:tauApart}), but also various asymptotic expressions.  For example, the asymptotic form of Equation~\eqref{eq:gamma} for values of $ka$ such that $\Delta' \delta_\eta\sim\Lambda\ll 1$ provides the ``FKR'' growth rate, $\gamma_{\rm FKR}$ \citep{furth1963finite}; values of $ka$ such that $\Delta'\delta_\eta\gg 1$ (i.e., $\Lambda\rightarrow 1$) provide the ``Coppi'' growth rate, $\gamma_{\rm Coppi}$ \citep{coppi1966resistive}. These limits are discussed, for example, in  \citet{boldyrev2018calculations}. To adapt these expressions for a poorly ionized plasma, we replace $v_{\rm A,i} \rightarrow v'_{\rm A,i}$ in Equation~\eqref{eq:gamma} and approximate the stability index using 
\begin{equation}\label{eq:Deltalim}
    \Delta'(k) \approx \frac{1}{ka^2}.
\end{equation}
This index is the small-$ka$ limit of the tearing stability index for both the classic \cite{harris1962plasma} equilibrium and, as we will show in Section~\ref{sec:pisp}, the poorly ionized CS (Equation~\eqref{eq:dpval}, with $a = x_1$). The resulting asymptotic expressions are used in \S\ref{sec:onset} to describe cases in which the plasma densities and collision frequencies can be assumed to be almost constant throughout the CS, and the magnetic field can be assumed to have a characteristic strength given by its value far from the sheet. Accordingly, in the remainder of this subsection we adorn values of densities, Alfv\'{e}n speeds, and collision frequencies  with the subscript ``0''.  
When ionization equilibrium holds, the effective Alfv\'en speed (Equation~\eqref{eq:sub}) may then be written as
\begin{equation}
 v'_{\rm A,i0} 
 = v_{\rm A,i0}\left( 1+ \frac{\rho_{\rm n0}}{\rho_{\rm i0}} \frac{\nu_{\rm ni0}+\xi}{\gamma+\nu_{\rm ni0}+\xi} \right)^{-1/2}
 .
\end{equation}
This expression reduces further in the strong-coupling approximation, for which $\xi\ll \nu_{\rm ni0}$:
\begin{equation}\label{eq:subsimp}
    v_{\rm A,i0}' \approx v_{\rm A,i0}\left[ 1+ \frac{\rho_{\rm n0}}{\rho_{\rm i0}} \left(1 + \frac{\gamma}{\nu_{\rm ni0}}\right)^{-1} \right]^{-1/2}.
\end{equation}
Let us also assume that the plasma is sufficiently poorly ionized and sufficiently unstable that $1\ll\gamma/\nu_{\rm ni0}\ll\rho_{\rm n0}/\rho_{\rm i0}$. Then we can simplify Equation~\eqref{eq:subsimp} all the way to 
\begin{equation}
\label{eq:vaiapprox}
    v'_{\rm A,i0}  \approx  v_{\rm A,n0}\left( \frac{\gamma}{\nu_{\rm ni0}} \right)^{1/2} ,
\end{equation}
in which case it becomes beneficial to introduce
\begin{equation}
   \tau_{\rm A,n0} \equiv (kv_{\rm A,n0})^{-1}
\end{equation}
as the characteristic timescale for an (undamped) Alfv\'en wave propagating in the ion--neutral fluid at speed $v_{\rm A,n0}$ to cross a fluctuation having wavenumber $k$. The poorly ionized version of the FKR growth rate, good for $S_{a_0}^{-1/6}\ll ka < 1$, is then
\begin{subequations}\label{eq:FKR}
\begin{align}
    \gamma_{\rm FKR} &\sim \tau_{\rm A,n0}^{1/2} \tau_\eta^{-3/4} \nu_{\rm ni0}^{-1/4} \, \frac{v_{\rm A,n0}}{a} \\*
    \mbox{} &= (ka)^{-1/2} \left(S^3_a \frac{a}{a_0}\right)^{-1/4} \frac{v_{\rm A,n0}}{a} .
\end{align}
\end{subequations}
This matches equation~(18) of \citet{zweibel1989magnetic} when $\Delta'$ is given by Equation~\eqref{eq:Deltalim}. In the Coppi limit, $ka \ll S_{a_0}^{-1/6}$ and we find that
\begin{subequations}\label{eq:Coppi}
\begin{align}
    \gamma_{\rm Coppi} &\sim \tau_{\rm A,n0}^{-1} \tau_\eta^{-1/2} \nu_{\rm ni0}^{-1/2} \\*
    \mbox{} & = ka  \left(S_a\frac{a}{a_0}\right)^{-1/2} \frac{v_{\rm A,n0}}{a}  . 
\end{align}
\end{subequations}
Because $\gamma_{\rm Coppi}\propto k$ while $\gamma_{\rm FKR}\propto k^{-1/2}$, we can obtain an approximate expression for the maximum tearing growth rate by setting Equations~\eqref{eq:FKR} and \eqref{eq:Coppi} equal to one another and finding the $k$ at which they intersect. At this maximally growing wavenumber, the growth rate is 
\begin{equation}
\label{eq:biggam}
    \gamma_{\rm t} \sim \left(\frac{v_{\rm A,n0}}{a\tau_\eta}\right)^{2/3} \nu^{-1/3}_{\rm ni0} = \left(S^2_a \frac{a}{a_0}\right)^{-1/3} \frac{v_{\rm A,n0}}{a}. 
\end{equation}
This matches equation~(14) of \citet{pucci2020tearing} for the fastest-growing tearing mode in the ``intermediate regime'' of $1\ll\gamma/\nu_{\rm ni0}\ll \rho_{\rm n0}/\rho_{\rm i0}$. The corresponding inner resistive layer thickness, defined in~\eqref{eq:reslayer}, satisfies $\delta_\eta/a \sim  S^{-1/6}_{a_0}$. 

By way of comparison, the maximum tearing growth rate in the regime where the ions and neutrals are fully decoupled from one another (i.e., $\gamma/\nu_{\rm ni0}\gg \rho_{\rm n0}/\rho_{\rm i0}$, such that $v'_{\rm A,i0}\approx v_{\rm A,i0}$) is given by
\begin{equation}
\label{eq:smallagam}
\gamma_{\rm t} \sim S^{-1/2}_{a} \left(\frac{\rho_{\rm n0}}{\rho_{\rm i0}}\right)^{1/4} \frac{v_{\rm A,n0}}{a}  .
\end{equation}
This approximate expression requires that the CS width satisfies $a/a_0\ll (\rho_{\rm i0}/\rho_{\rm n0})^{1/2} S^{-1/3}_{a_0}$.

\section{Multi-Stage CS Formation}
\label{sec:natform}
\begin{figure*}[ht!]
\centering
\includegraphics[width=5.8cm]{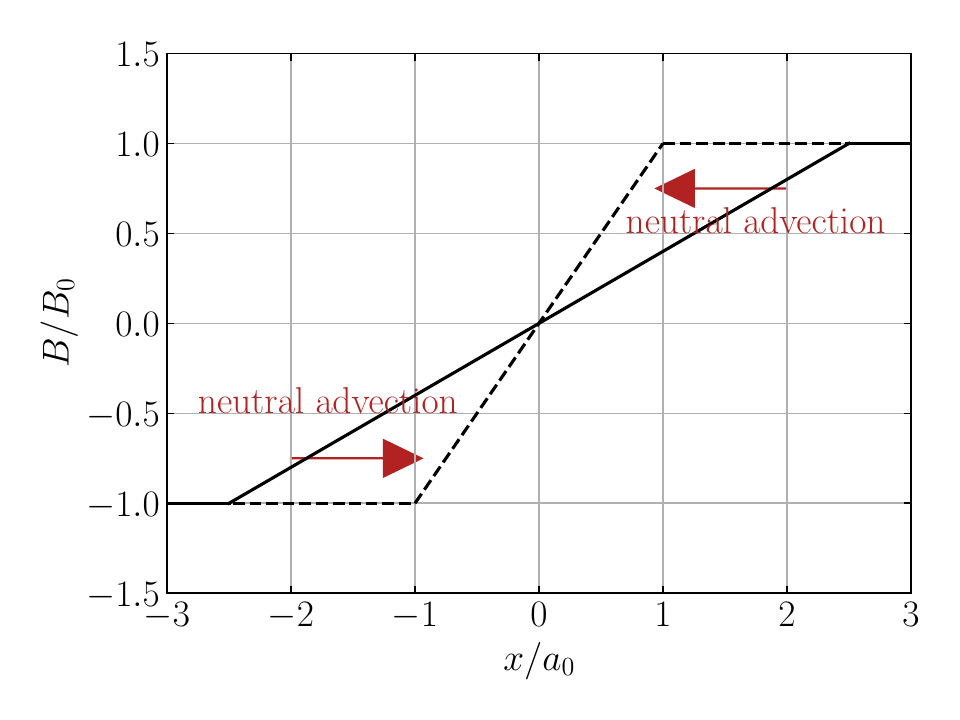}
\includegraphics[width=5.8cm]{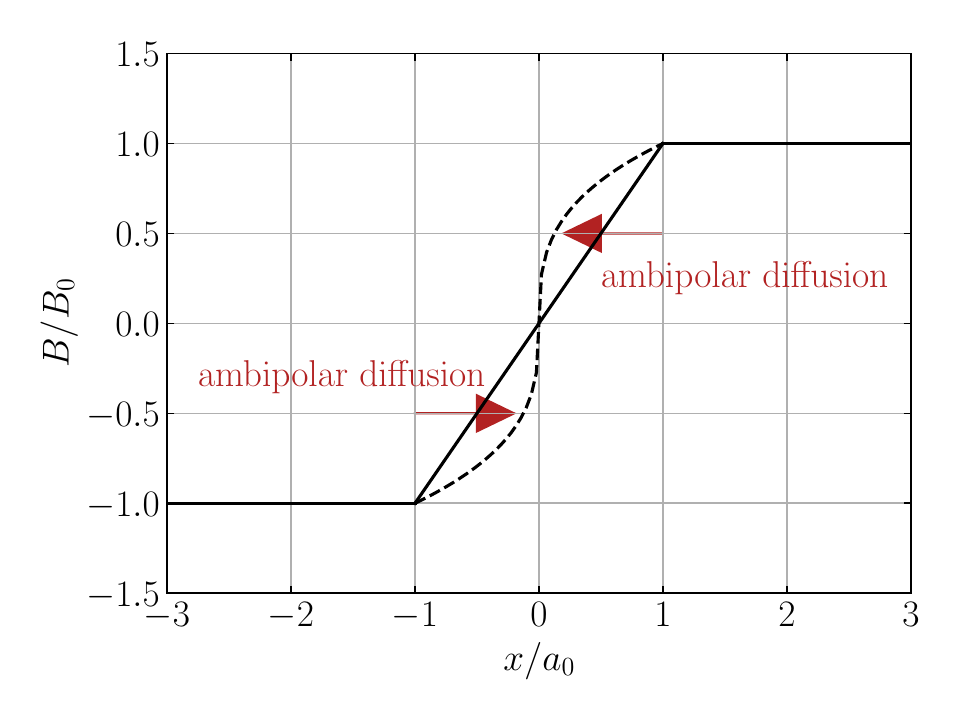}
\includegraphics[width=5.8cm]{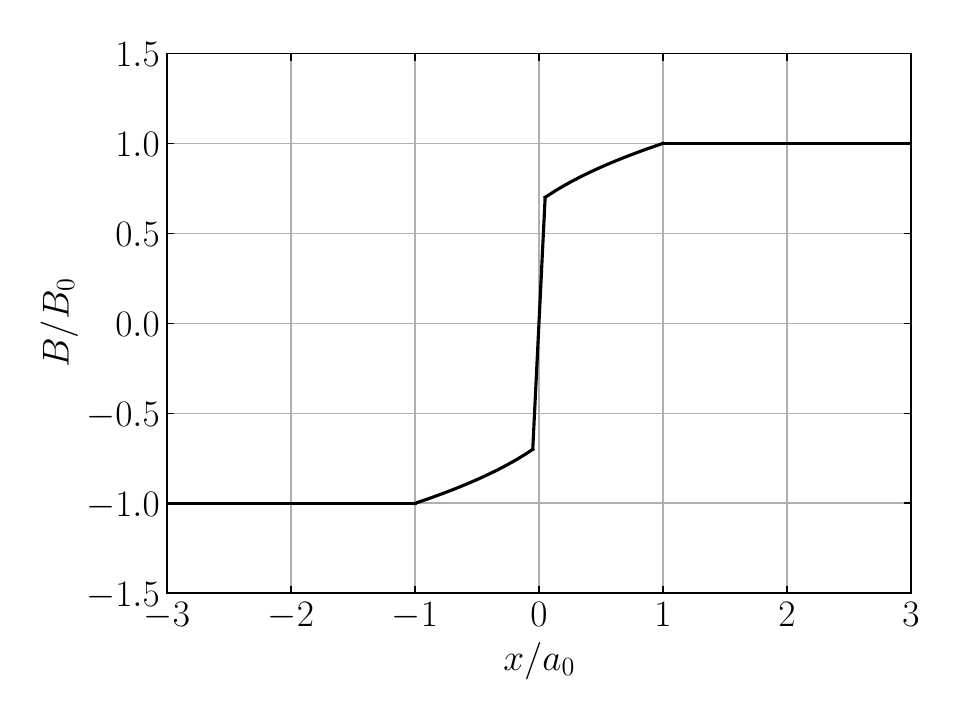}
      \caption{Schematic diagrams of the three stages of CS formation, from left to right: (1) an initial neutral-advection driven stage, (2) an ambipolar diffusion stage, and (3) a final steady-state CS.}
\label{fig:formation}
\end{figure*}
In this Section, we describe the evolution of the profile in Equation~\eqref{eq:maginit} in the absence of the tearing instability. This description will allow us to understand when in the formation process tearing onsets. The evolution of the profile comprises three stages: a stage in which neutral advection dominates CS evolution, a stage in which ambipolar diffusion dominates the evolution, and a final steady-state stage. These stages are diagrammed schematically in Figure~\ref{fig:formation}; we describe each stage, and note relevant characteristics about it, in the Sections that follow.

\subsection{Neutral-Advection-Dominated Formation}
\label{sec:formsheet}
We begin by considering the first stage, in which the CS reverses over a large length scale $a(t)$. 
By considering the relative sizes of the terms in Equation~\eqref{eq:Bstrong} pertaining to neutral advection and ambipolar diffusion, we see that, on length scales above $a_0$, changes in the magnetic field are caused primarily by neutral advection. Thus, initially, the CS is formed by flows in the coupled neutral fluid, as depicted in the leftmost panel of Figure~\ref{fig:formation}. How precisely this stage occurs and how long it lasts depend on the nature of those flows, which vary from context to context.  However, we can describe the time-dependent rate of formation during this stage as 
\begin{equation}
\label{eq:tauNA}
    \tau_{\rm NA}^{-1} \sim \frac{v_{\rm A,n0}}{a(t)}.
\end{equation}
This approximate expression will be helpful in Section~\ref{sec:onset} in determining if tearing can onset during this stage.

\subsection{Ambipolar-Diffusion-Dominated Formation}
\label{sec:ambipolardiff}
Once motions in the neutral fluid thin the CS enough that it varies on a length scale $a(t)\lesssim a_0$, the ambipolar diffusion term in Equation~\eqref{eq:Bstrong} dominates, and the second stage of CS formation begins. As in Section~\ref{sec:pstate}, we assume that  neutral advection constantly supplies magnetic flux to the CS, such that the magnetic field is pinned to a constant value at the edge of the region in which ambipolar diffusion is occurring,
\begin{equation}
\label{eq:edgeB}
    B_y(\pm a_0, t ) = \pm B_0.
\end{equation}
Likewise, we take the plasma at the edge of the ambipolar region to have constant ionized and neutral fluid densities, $\rho_{\rm i0}$ and $\rho_{\rm n0}$ respectively, with their ratio determined by ionization equilibrium, Equation~\eqref{eq:ionequil}. Because ambipolar diffusion is nonlinearly dependent upon the magnetic-field strength, it attempts to sharpen the CS towards the singular current configuration given by Equation~\eqref{eq:generalform2} \citep{brandenburg1994formation,brandenburg1995effects}, as described in Section~\ref{sec:strongcoupling} and as depicted schematically in the center panel of Figure~\ref{fig:formation}. 

Initially, the rate of CS formation increases with time, reflecting that the ambipolar-diffusion rate, which can be estimated using Equation~\eqref{eq:bev} as
\begin{equation}
\label{eq:tad}
    \tau_{\rm AD}^{-1} \sim \frac{v_{\rm A,n0}^2}{a^2 \nu_{\rm ni0}} = \nu_{\rm ni0} \left(\frac{a_0}{a}\right)^2,
\end{equation}
increases with decreasing CS width $a$. Eventually, after the magnetic field has diffused nearly all of the way to the center of the null,  this sharpening process is curbed by contributions from resistivity and finite ionized fluid pressure \citep{brandenburg1995effects}. These cause the strong-coupling assumption to break down near the center of the sheet.\footnote{In Section~\ref{sec:pisp}, we show that the  pressure gradient of the ionized fluid becomes important when the CS width $a$ approaches the value $x_1$ that characterizes the effective width of a poorly ionized Sweet--Parker CS, Equation~\eqref{eq:x1}. This width is much smaller than the width of the ambipolar diffusion region surrounding the magnetic null, i.e., $x_1/a_0 \ll 1$.} The initial width of the magnetic null is $a =a_0$, so that we can use Equation~\eqref{eq:tad} to estimate the duration of the second stage of CS formation, $t_{\rm form}$, as
\begin{equation}
\label{eq:tform}
t_{\rm form} \sim \nu_{\rm ni0}^{-1} ;
\end{equation}
i.e., the time for the CS to steepen nonlinearly by ambipolar diffusion is comparable to the initial neutral--ion collision time.

\subsection{Steady-State CS}
\label{sec:pisp}

After the second stage of CS formation occurs, the CS enters its final stage, a steady-state CS. This stage can be thought of as an analog of the Sweet--Parker CS in a fully ionized plasma. This stage is depicted in the final panel of Figure~\ref{fig:formation}. Here, we describe the structure of this sheet.

Close to the singularity at $x=0$, Equation~\eqref{eq:generalform} is no longer valid. In this region, the current singularity suggested by the ambipolar diffusion term is resolved by resistivity and finite ionized fluid pressure \citep{brandenburg1995effects}, and we cannot apply the strong-coupling approximation, Equation~\eqref{eq:strongcoupling}.  In steady state, we have from Equation~\eqref{eq:induction} that near the magnetic null at $x=0$,
\begin{equation}
\label{eq:nullcond}
    0 \approx \grad \btimes ( \bb{v}_{\rm i} \btimes \bb{B} ) - \eta \nabla^2 \bb{B} \approx - \bb{\hat{y}} \,\eta \partial_{x}^2 B_y,
\end{equation}
where the final approximate equality comes from the recognition that in steady state near a magnetic null, the electric field (determined from Equation~\eqref{eq:Ohm}) should remain constant while the magnetic field and $\bb{\hat{x}}$-directed ionized fluid velocity go to zero, such that the resistive term dominates \citep{brandenburg1995effects}. 
Condition~\eqref{eq:nullcond} implies that the magnetic-field strength should be linear near the center of the CS. Far from the center of the CS, the functional form enforced by ambipolar diffusion, described by Equation~\eqref{eq:generalform}, continues to hold.
Altogether then, the magnetic field within the ambipolar-diffusion-dominated region can be described by 
\begin{align}
\label{eq:btot}
    &B_{y} (x) \nonumber\\*
    \mbox{} &=
    \begin{cases}
           \left[ \dfrac{B_0^3 (x + x_1 ) -B_1^3 (a_0+x) }{a_0-x_1} \right]^{1/3}, &-a_0 \le x\le -x_1 \vspace{1ex}\\
            \dfrac{B_1 x}{x_1}, & -x_1< x< x_1 \vspace{1ex}\\
            \left[ \dfrac{B_0^3 \left(x-x_1 \right) + B_1^3 \left(a_0-x \right) }{a_0-x_1} \right]^{1/3}, & x_1 \le x \le a_0,
    \end{cases}
\end{align}
where $x_1$ is the point where the strong-coupling assumption breaks and $B_1$ is the magnetic-field strength at that point. In the region where the magnetic field is linear, the magnetic pressure gradient is supported by a sharp gradient in the pressure of the ionized fluid. This magnetic configuration is shown schematically in the rightmost panel of Figure~\ref{fig:formation}.

Precise values for the parameters in Equation~\eqref{eq:btot} must be found numerically. However, we can motivate scaling laws for how they depend on the plasma properties by adopting physical arguments supported by a more detailed study of a one-dimensional version of our setup \citep{heitsch2003fast}. These arguments also provide scaling laws for the ionized fluid inflow velocity to the sheet, $v_{\rm i,AD}$, which transports magnetic flux inwards and supports the magnetic profile in the outer region; the characteristic velocity in the linear region $-x_1<x<x_1$, $v_{\rm i,\eta}$ (which is smaller than $v_{\rm i,AD}$); and the ionized fluid over-density at the center of the sheet, $\rho_{\rm i,CS}$. 

First, the continuity equation for the ionized fluid, Equation~\eqref{eq:contn}, implies that plasma moving inside the inner linear-magnetic-field region recombines, yielding the following balance:\footnote{The balance given by  Equation~\eqref{eq:firstbalance} only holds in plasmas with $\alpha$ large enough that the recombination term in Equation~\eqref{eq:conti} is more important than the portion of $\grad \bcdot ( \rho_{\rm i} \bb{v}_{\rm i} )$ due to outflows along the CS. This occurs when $\alpha \gg (v_{\rm out}/L_{\rm CS})(C_{\rm i}/B_0)^2$, with $v_{\rm out}$ the outflow speed and $L_{\rm CS}$ the CS length. If this condition does not hold, then the sheet becomes a typical Sweet--Parker CS.}
\begin{equation}\label{eq:firstbalance}
    \frac{\rho_{\rm i,CS} v_{\rm i,\eta }}{x_1} \sim \alpha \rho_{\rm i,CS}^2.
\end{equation}
Second, the induction equation~\eqref{eq:induction} states that the magnetic field in the inner region is diffused away by resistivity, giving
\begin{equation}
    x_1 \sim \frac{\eta}{v_{\rm i,\eta}}.
\end{equation}
Third, approximate pressure balance throughout the CS provides
\begin{equation}
    \rho_{\rm i,CS} C_{\rm i}^2 + \rho_{\rm n,CS} C_{\rm n}^2  \sim \frac{B_1^2}{2} + \rho_{\rm n1} C_{\rm n}^2 \sim \frac{B_0^2}{2} + \rho_{\rm n0} C_{\rm n}^2,
\end{equation}
where $\rho_{\rm n,CS}$ is the neutral fluid density at the center of the CS, $\rho_{\rm n1}$ is the neutral fluid density at $x_1$, and we have neglected the contribution of the ionized fluid pressure at $x_1$ and far from the CS.  To simplify this expression further, we approximate that the neutral fluid pressure is approximately constant throughout the CS, so that it is primarily the gradient in the ionized fluid pressure (governed by recombination at the center of the CS) that balances the magnetic pressure gradient. As a result,
\begin{equation}
\label{eq:rhoics}
  \rho_{\rm i,CS} \sim \frac{B^2_1}{C^2_{\rm i}}
\end{equation}
and
\begin{equation}
\label{eq:b1}
B_1 \sim B_0.
\end{equation}
The range of validity of this assumption is checked {\em a posteriori}. From these balances, Equations~\eqref{eq:firstbalance}--\eqref{eq:b1}, we find the following scaling laws for the CS width\footnote{An astute reader might wonder about our theory's applicability in the asymptotic limit $\alpha \rightarrow 0$.  However, our theory only attempts to treat plasmas in which the strong-coupling approximation is broken at a length scale below the ambipolar-diffusion length scale. Thus it is only good for $x_1 \ll a_0$. The expression~\eqref{eq:x1} shows that this is a large-$\alpha$ limit.}
\begin{equation}
\label{eq:x1}
    x_1 \sim \frac{C_{\rm i}}{B_0} \sqrt{\frac{\eta}{\alpha}}
\end{equation}
and the reconnecting ionized fluid velocity
\begin{equation}
\label{eq:vini}
    v_{\rm i, \eta } \sim \frac{B_0}{C_{\rm i}} \sqrt{\alpha\eta} .
\end{equation}
From mass flux continuity we can estimate 
\begin{equation}
\rho_{\rm i0}  v_{\rm i,AD} \sim \rho_{\rm i,CS} v_{\rm i,\eta} ,  
\end{equation}
or, using Equations~\eqref{eq:rhoics}, \eqref{eq:b1}, and \eqref{eq:vini},
\begin{equation}
\label{eq:viad}
v_{\rm i,AD} \sim v_{\rm i,\eta} \, \frac{v^2_{\rm A,i0}}{C_{\rm i}^2} \sim \frac{B_0}{C_{\rm i}} \frac{\sqrt{\alpha\eta}}{\beta_{\rm i0}}  .
\end{equation}
Under these scalings, a larger recombination coefficient implies a larger ionized fluid inflow velocity.

To determine the range of validity of these scalings, we can use Equation~\eqref{eq:viad} to check the relation \eqref{eq:b1}. In particular, in the case when $B_1$ differs noticeably from $B_0$, we can relate the inflow velocity to $B_1$ by using Equation~\eqref{eq:btot} to evaluate $\lim \limits_{x\rightarrow x_1^{+}} \grad\left(B^2/2\right)$ and then applying the strong-coupling relation~\eqref{eq:strongcoupling} with $|\bb{v}_{\rm i} - \bb{v}_{\rm n}| \sim |\bb{v}_{\rm i}|$. This gives $B_1/B_0 \sim B_0^2/(\rho_{\rm n0} v_{\rm A,n0} v_{\rm i,AD})$. Using Equation~\eqref{eq:viad}, we find that $B_1/B_0 \ll 1$ when $C_{\rm i}\beta_{\rm i0}/\sqrt{\alpha\eta\rho_{\rm n0}} \ll 1$, and our assumption that $B_1 \sim B_0$ then fails.  This case corresponds to the ``$Z \ll 1$''  case in~\cite{heitsch2003fast}, in which scalings different from those above must be used. Thus, our results are good only for 
\begin{equation}
\label{eq:validcond}
    \mathcal{Z} \equiv \frac{C_{\rm i}\beta_{\rm i0}}{\sqrt{\alpha\eta\rho_{\rm n0}}} \gtrsim 1 ;
\end{equation}
i.e., the recombination rate must be sufficiently small to allow the ionized fluid pressure in the inner region $x_1$ to grow large enough to support the CS's magnetic pressure. This criterion is satisfied in all of our numerical simulations (\S\ref{sec:simulation}), where $\alpha$ must be small so that $x_1$ is a resolvable length, but it breaks down in systems with low ionization fractions (such as star-forming molecular clouds; see Appendix~\ref{sec:typval}). In this case (i.e., $\mathcal{Z}\ll 1$), our expression for $x_1$ (Equation~\eqref{eq:x1}) simply acquires a multiplicative factor of $\mathcal{Z}^{-1/4}$ \citep{heitsch2003fast}.

In the next section (\S\ref{sec:initonset}), these expressions are used to identify the tearing mode that disrupts the CS. To do this, it is also necessary to have a quantitative expression for the tearing-mode stability index $\Delta'$ in Equation~\eqref{eq:gamma} for the profile~\eqref{eq:btot}.
This is derived in Appendix~\ref{sec:deltaprime}, with the result being that
\begin{equation}
\label{eq:dpval}
    \Delta'(k) = k\left[-2 + \dfrac{4}{1 + (2 k x_1 -1 )\exp(2 k x_1)} \right] .
\end{equation}
For small $kx_1$, this quantity approaches $\Delta' \approx (kx_1^2)^{-1}$, which agrees with the trend found numerically by \cite{brandenburg1995effects}. Note that this long-wavelength limit matches the tearing-mode stability index corresponding to a \citet{harris1962plasma} equilibrium having width $x_1$.

\section{Tearing Onset in the Forming CS}
\label{sec:onset}
The goal of the paper is to determine when and how in the formation process outlined in Section~\ref{sec:natform} tearing onsets. We conduct this task in this Section, starting by determining at what time tearing onsets and concluding by determining the wavelength of the mode that disrupts the sheet.

\subsection{Time of Tearing Onset}
Tearing onset in time-dependent systems has been considered in several recent works \citep{pucci2013reconnection,uzdensky2016magnetic,comisso2016general,huang2017plasmoid,tolman2018development}. A rough consensus of these works is as follows. 

First, a necessary requirement for tearing growth is that the tearing stability index $\Delta'(k)$ be positive for a wavenumber $k$ that fits within the length of the CS.  The value of $\Delta'$ increases as $k$ and the width decrease. The minimum possible value of $k$ is $k = 2\pi/L_{\rm CS}$ with $L_{\rm CS}$ the CS length.  We assume that our CS is sufficiently long that a $k$ small enough for $\Delta'$ to be positive exists. 

Second, the growth rate of the unstable tearing mode, given implicitly by Equation~\eqref{eq:gamma} and approximately by Equation~\eqref{eq:biggam}, must also be large enough to exceed significantly the rate of CS formation, allowing for an asymptotic separation between the evolution of the tearing perturbation and the evolution of the background CS \citep{tolman2018development}. Let us consider when and if these conditions are met during the formation process outlined in Section~\ref{sec:natform}.

In this consideration, we require an expression for the tearing stability index $\Delta'$. The form of this index depends on the functional form of the magnetic field, and the evolving CS experiences a variety of magnetic configurations. For our analysis, we choose to use the stability index
\begin{equation}
\label{eq:dpapprox}
    \Delta'(k) \approx \frac{1}{k a^2},
\end{equation}
which was used to derive the approximate expression~\eqref{eq:biggam} for the maximum tearing  growth rate.
This index is the small-$ka$ limit of the tearing stability index for both the classic \cite{harris1962plasma} equilibrium and the poorly ionized CS (Equation~\eqref{eq:dpval}, with $a = x_1$).\footnote{For values of $kx_1$ greater than  ${\simeq}0.64$, the value of $\Delta'$ given by Equation~\eqref{eq:dpval} is negative and no instability exists. Thus, the small-$ka$ limit is likely a good approximation for any CS of the poorly ionized form that is unstable.} Thus, Equation~\eqref{eq:dpapprox} is expected to be sufficient to describe overall trends occurring in a generic narrowing magnetic configuration.

During the neutral-advection-dominated formation stage discussed in Section~\ref{sec:formsheet}, we have from Equations~\eqref{eq:biggam} and \eqref{eq:tauNA} that
\begin{equation}
    \gamma_t \tau_{\rm NA} \sim \frac{\eta^{2/3}}{a (t) v_{\rm A,n0}^{1/3} \nu_{\rm ni0}^{1/3}} \lesssim \frac{\eta^{2/3} \nu_{\rm ni0}^{2/3}}{v_{\rm A,n0}^{4/3}}  = S_{a_0}^{-2/3},
\end{equation}
where in the final inequality we have recognized that the minimum value of $a(t)$ during this stage is $a_0$. The value of  $ S_{a_0}^{-2/3} $ is  small in high-Lundquist number plasmas, showing that the tearing growth rate is always smaller than the CS formation rate. As a result, tearing cannot onset during this stage.

Next we consider if an asymptotic separation of timescales between the formation rate and the tearing growth rate can occur during the ambipolar-diffusion stage discussed in Section~\ref{sec:ambipolardiff}. We  compare the fastest tearing growth rate (Equation~\eqref{eq:biggam}) to the rate of CS formation via ambipolar diffusion, Equation~\eqref{eq:tad}:
\begin{equation}
    \gamma_{\rm t}\tau_{\rm AD} \sim \frac{\eta^{2/3} \nu_{\rm ni0}^{2/3}}{v_{\rm A,n0}^{4/3}},
\end{equation}
which is again a small number that remains nearly constant throughout CS formation. Asymptotic separation between the tearing growth rate and CS formation rate cannot be achieved during this second stage.\footnote{Very late in the formation of the CS, the CS width may become sufficiently small that Equation~\eqref{eq:smallagam} must be used for the maximal tearing growth rate. In this case, $\gamma_t \tau_{AD}$ is still small and tearing cannot onset.}

However, in the third stage identified in Section~\ref{sec:natform}, when the strong-coupling approximation breaks down and the CS settles into a steady state with a steep inner profile, tearing can easily onset. At this time, given by Equation~\eqref{eq:tform}, the CS formation rate slows to zero, and the unstable tearing modes can grow. Thus, we predict that the tearing onset time is given by
\begin{equation}
\label{eq:tonset}
t_{\rm onset} \sim t_{ \rm form} \sim \nu_{\rm ni0}^{-1}.
\end{equation}
We conclude that tearing onsets at the end of the second stage, while the breaking of strong coupling arrests CS formation. 

\subsection{Dimensionless Variables}
We have identified in Equation~\eqref{eq:a0} the characteristic length scale of the problem, $a_0$, and in Equation~\eqref{eq:tonset} the characteristic timescale of the problem, $\nu_{\rm ni0}^{-1}$.  Henceforth, we refer to quantities normalized to these values and to the background plasma parameters, with normalized quantities being adorned with a tilde. That is, we replace
\begin{subequations}
\begin{equation}
    \pD{t}{} \rightarrow \nu_{\rm ni0} \pD{\tilde{t}}{} \qquad{\rm and}\qquad \grad\rightarrow \frac{1}{a_0} \tilde{\grad} , \tag{\theequation {\it a},{\it b}}
\end{equation}
and introduce the dimensionless variables
\begin{gather}
    \tilde{\rho}_{\rm i} \equiv \frac{\rho_{\rm i}}{\rho_{\rm n0}} , \quad\tilde{\rho}_{\rm n}\equiv\frac{\rho_{\rm n}}{\rho_{\rm n0}} ,  \tag{\theequation {\it a},{\it b}}\\*
    \tilde{P}_{\rm i} \equiv \frac{P_{\rm i}}{\rho_{\rm n0}a^2_0 \nu^2_{\rm ni0}} , \quad \tilde{P}_{\rm n}\equiv \frac{P_{\rm n}}{\rho_{\rm n0}a^2_0\nu^2_{\rm ni0}} ,  \tag{\theequation {\it c},{\it d}}\\*
    \tilde{v} \equiv \frac{v}{a_0\nu_{\rm ni0}},\quad \tilde{B} \equiv \frac{B}{a_0\nu_{\rm ni0}\sqrt{\rho_{\rm n0}}} ,  \tag{\theequation {\it e},{\it f}}\\*
    \tilde{\xi}\equiv\frac{\xi}{\nu_{\rm ni0}}, \quad \tilde{\alpha}\equiv\frac{\alpha\rho_{\rm n0}}{\nu_{\rm ni0}}, \quad \tilde{\eta}\equiv\frac{\eta}{a^2_0\nu_{\rm ni0}} . \tag{\theequation {\it g}}
\end{gather}
\end{subequations}

Note that $\tilde{\eta}$ is equivalent to the inverse Lundquist number defined using $a_0$, $S_{a_0}^{-1}$. The dimensionless neutral--ion collision frequency,  
\begin{equation}
    \tilde{\nu}_{\rm ni0} \equiv \frac{\nu_{\rm ni} }{\nu_{\rm ni0} }= \tilde{\rho_{\rm i}} \, \sqrt{\frac{\tilde{\alpha}}{\tilde{\xi}}},
\end{equation}
is a function of the normalized ionized fluid density, recombination coefficient, and  ionization coefficient. The cross section and mass that enter in Equation~\eqref{eq:nuni} affect the lengthscales and timescales of the problem but no other element of the physics. 
When normalized, the ionized momentum equation~\eqref{eq:momi} becomes
\begin{align}\label{eq:ionpn}
    \pD{\tilde{t}}{(\tilde{\rho}_{\rm i}\tilde{\bb{v}}_{\rm i})} &+ \tilde{\grad}\bcdot(\tilde{\rho}_{\rm i}\tilde{\bb{v}}_{\rm i}\tilde{\bb{v}}_{\rm i}) = -\tilde{\grad} \tilde{P}_{\rm i} + (\tilde{\grad}\btimes\tilde{\bb{B}})\btimes\tilde{\bb{B}} \nonumber\\*
    \mbox{} &-\tilde{\nu}_{\rm ni} \tilde{\rho}_{\rm n}(\tilde{\bb{v}}_{\rm i}-\tilde{\bb{v}}_{\rm n}) + \tilde{\xi}\tilde{\rho}_{\rm n}\tilde{\bb{v}}_{\rm n} - \tilde{\alpha} \tilde{\rho}^2_{\rm i}\tilde{\bb{v}}_{\rm i},
\end{align}
with similar forms holding for the other equations we solve.

\subsection{Tearing Mode That Disrupts the CS}
\label{sec:initonset}
Let us now consider how tearing onsets when the strong-coupling approximation breaks. The strong coupling approximation does not break instantaneously; after the ambipolar diffusion stage, CS formation slows down continuously until the CS reaches its final state, in which the magnetic configuration does not change. Tearing onsets at some point during this slowing-down process, before CS formation has ceased completely.  
However, expressions for the fully formed CS parameters, given in Section~\ref{sec:pisp}, are sufficient to describe the CS at the time when tearing onsets, and the onset time is given approximately by the formation time, Equation~\eqref{eq:tform}. 

\begin{figure}
\centering
  \includegraphics[width=\columnwidth]{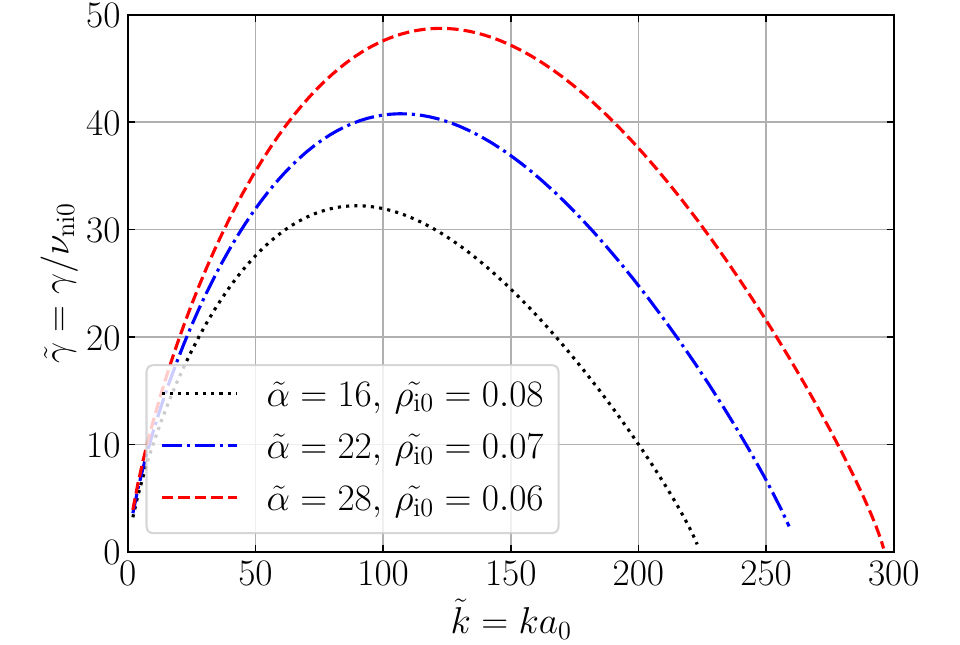}%
  \caption{Linear tearing growth rate $\tilde{\gamma}$ vs.~wavenumber $\tilde{k}$ in a poorly ionized CS for different recombination coefficients $\tilde{\alpha}$. Each curve is calculated by evaluating Equation~\eqref{eq:gamma}, with $\Delta'(k)$ given by Equation~\eqref{eq:dpval}. In all cases,  $\tilde{\xi} = 0.1$, $\tilde{\eta} = 10^{-4}$, and $\tilde{C}_{\rm i} = \tilde{C}_{\rm n} = 1$. The proportionality coefficient for $\tilde{x}_1$ in Equation~\eqref{eq:x1} and the values for ionized density and magnetic field are set using our simulation results (see Equations~\eqref{eq:x131}--\eqref{eq:B131}). 
  }
\label{fig:grlin}
\end{figure}

When tearing onsets, a wide range of modes could be destabilized. These modes have growth rates determined by Equation~\eqref{eq:gamma} modified according to Equation~\eqref{eq:tauApart}, with the tearing stability index given by Equation~\eqref{eq:dpval}.\footnote{Technically, tearing instability occurs in a background that includes the flows estimated by Equations~\eqref{eq:vini} and \eqref{eq:viad}. These flows do not influence tearing if the tearing growth rate is much larger than the flow shear rate \citep{tolman2018development}, i.e., 
\begin{equation}\label{eq:condition}
    \frac{\gamma_{\rm t}}{v_{ \rm flow}/a_{ \rm flow}} \gg 1,
\end{equation}
where $\gamma_{\rm t}$ is the maximum growth rate and $a_{\rm flow}$ is the length scale of the flow speed $v_{\rm flow}$.  The flow shear is highest in the inner region of the CS, and so we may estimate the denominator of Equation~\eqref{eq:condition} using Equations~\eqref{eq:x1} (i.e., $a_{\rm flow}\sim x_1$) and \eqref{eq:vini} (i.e., $v_{\rm flow}\sim v_{\rm i,\eta}$). Then we find that, for the tearing growth rate given by Equation~\eqref{eq:biggam}, condition~\eqref{eq:condition} holds if $\eta^{-1/3}  v_{\rm A,n}^{2/3} \nu_{\rm ni}^{-1/3} \gg 1$. This condition holds in all of the high-Lundquist-number plasmas that we study.} (Full dispersion relations, rather than asymptotic expressions, allow a deeper understanding of this stage.) A plot of such growth rates for three different recombination parameters $\tilde{\alpha}$ is shown in Figure~\ref{fig:grlin}. The mode with the largest growth rate grows fastest, and quickly overwhelms any other modes that might grow. Thus, for a given set of plasma parameters, we can predict the mode that dominates linear onset by finding the value of $\tilde{k}$ that maximizes Equation~\eqref{eq:gamma} modified according to Equation~\eqref{eq:tauApart}, with $\tilde{a}$ set to Equation~\eqref{eq:x1} and the ionized fluid Alfv\'{e}n speed defined using  Equations~\eqref{eq:rhoics} and \eqref{eq:b1}. 

\begin{figure}
\centering
  \includegraphics[width=\columnwidth]{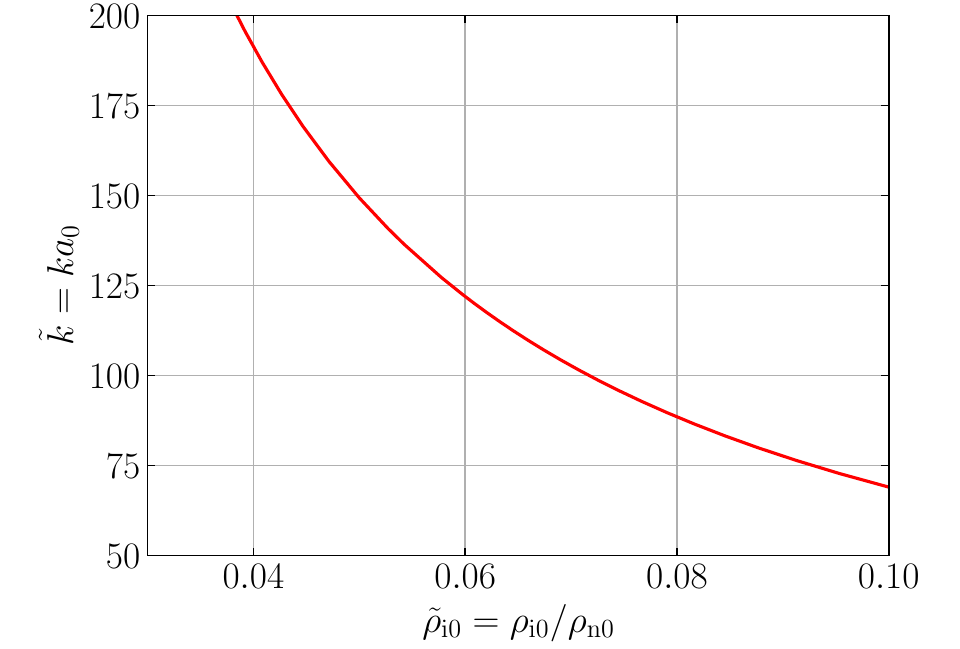}%
  \caption{Wavenumber $\tilde{k}$ of the tearing mode predicted to  dominate linear onset in a poorly ionized CS for different $\tilde{\alpha}$ and, therefore, different equilibrium ionization fractions $\tilde{\rho}_{\rm i0}$. Other parameters are as in Figure~\ref{fig:grlin}.
  }
\label{fig:konset}
\end{figure}

The resulting values of $\tilde{k}$ for a set of ionization fractions are shown in Figure~\ref{fig:konset}.  This figure illustrates a central finding of this paper: as the plasma recombination rate increases and ionization fraction decreases, the wavelength of the mode that dominates the linear stage becomes a decreasing fraction of $a_0$.

\section{Illustration and Verification by Direct Numerical Simulations}
\label{sec:simulation}
In this Section, we test the conclusions developed analytically in previous Sections of the paper numerically. Namely, we use nonlinear two-fluid numerical simulations to demonstrate that ambipolar diffusion can cause a CS to evolve from tearing stable to tearing unstable, that this process occurs in a time $\tilde{t} \sim 1$, and that the wavenumber that dominates the linear onset can be found as described in Section~\ref{sec:initonset}.  We also verify multiple aspects of the tearing physics developed in Appendix~\ref{sec:stationary}, including the tearing-mode eigenfunction and its growth rate. 

\subsection{Numerical Approach}

We perform a suite of simulations demonstrating CS formation, tearing onset, linear tearing growth, and the production of a chain of nonlinear plasmoids (studied further in~\cite{tolman23II}). Our numerical simulations employ the GPU-enabled astrophysical magnetohydrodynamic code suite {\tt AthenaK},
which allows the simultaneous, coupled simulation of magnetohydrodynamic and hydrodynamic species. This suite also implements novel implicit-explicit methods \citep{pareschi2005implicit} that enable the efficient treatment of the different timescales of these species, which has historically been an impediment to the study of reconnection in poorly ionized plasmas  \citep{zweibel2015ambipolar}. 

All of our simulations focus on the portion of the CS in which ambipolar diffusion is important, with the active mesh covering the range  $\tilde{x} \in [-1,1]$.  The length of the active mesh is $\tilde{y} \in [-2,2]$; the longer $\tilde{y}$ domain helps the plasmoids escape in the nonlinear portion of the simulation.  Following Equation~\eqref{eq:maginit}, the magnetic field in the active mesh lies primarily in the $\hat{y}$ direction.  However, we introduce a slight pinch in this field towards $\tilde{x} = 0$ to aid the escape of any plasmoids produced from the simulation domain.   Specifically, the initial magnetic configuration is chosen to reflect the magnetic-field structure between the boundary of two adjacent magnetic islands (depicted in Fig.~1 of \cite{huang2010scaling}), i.e., we adopt a vector potential given by 
\begin{equation}
\label{eq:vecpot}
    \tilde{\bb{A}}(\tilde{x},\tilde{y}) = -\frac{5}{\pi}   \tanh( \tilde{x}) \sin(\pi  \tilde{x}/5 ) \cos(\pi \tilde{y} / 9)\,\ez .
\end{equation}
The coefficients $5$ and $9$ are selected such that the magnetic-field strength at $(\tilde{x}, \tilde{y}) = (\pm 1, 0)$ is close to ${\pm}1$ and the pinch is sufficient to help plasmoids escape but is not too pronounced. 

The  neutral fluid pressure is taken to balance the magnetic pressure approximately, following the method of \cite{huang2010scaling}; the ionized fluid pressure is set such that the ionized fluid density is in ionization balance with the neutral fluid density. 
As in other simulations in the literature demonstrating sharpening of a CS caused by ambipolar diffusion \citep[e.g.,][]{brandenburg1995effects}, this is not a pressure-balanced equilibrium, but it very quickly settles into one: the ionized fluid starts to move, and then drags on the neutrals, such that the neutral fluid pressure begins to support the magnetic pressure via the drag term. We verify that our simulations approximately satisfy the strong coupling condition~\eqref{eq:strongcoupling} soon after the CS starts to evolve.

To introduce perturbations that can grow into tearing eigenmodes and to introduce asymmetry in the $\hat{y}$ direction (necessary to prevent the artificial lingering of plasmoids near the center of the sheet), the ionized fluid is seeded with Gaussian-random momentum perturbations in the $\hat{x}$ and $\hat{y}$ directions having amplitude $10^{-7}$.

\begin{table}
  \begin{center}
  \begin{tabular}{|c|c|}
 Parameter &Value  \\[3pt] \hline
$\tilde{\xi}$&  $0.10$  \\ 
$\tilde{\alpha}$&  $10., \, 16, \, 22, \, 28, \, 34, \, 40., \, 60.$  \\ 
$\tilde{\eta}$  & $10^{-4}$ \\
$\tilde{C}_{\rm i}, \, \tilde{C}_{\rm n}$  & $1.0$ \\
$\tilde{\rho}_{\rm n 0 }$ & $1.0$ \\
$\tilde{\rho}_{\rm i0} = (\tilde{\xi}/\tilde{\alpha})^{1/2} ~[10^{-2}]$  & $10., 7.9, \,6.7, \,6.0, \,5.4, \, 5.0, \, 4.1$ \\
$\beta_{\rm i0} = 2\tilde{\rho}_{\rm i0}\tilde{C}_{\rm i}^2 /\tilde{B}^2 ~[10^{-2}]$  & $20., \, 16, \, 13, \, 12, \, 11, \, 10., \, 8.2  $  \\
$\beta_{\rm n0} = 2  \tilde{\rho}_{\rm n0}\tilde{C}_{\rm n}^2/\tilde{B}^2$  & $2.0$ \\
$\mathcal{Z} = \tilde{C}_{\rm i}\beta_{\rm i0}/(\tilde{\alpha}\tilde{\eta}\tilde{\rho}_{\rm n0})^{1/2}$ &  $6.2, \, 4.0, \, 2.9, \, 2.3, \, 1.9, \, 1.6, \, 1.1$ \\
maximum SMR level & $6, \, 6, \, 7, \, 7, \, 7, \,7, \, 7$ \\
  \end{tabular}
  \caption{Initial parameters used in the two-fluid {\tt AthenaK} simulations.}
  \label{tab:params}
  \end{center}
\end{table}

Near the center of the CS, these simulations all show tearing behavior, which has very fine-scale structure that is only resolvable with high spatial resolution. Far from the CS, where no instability exists, these simulations exhibit only ambipolar diffusion, which results in relatively large-scale structure and therefore is resolvable with less resolution.  Accordingly, we run our simulations using static mesh refinement (SMR): far from the center of the CS, the resolution of the simulation is $256\times 512$, while the inner part of the CS is resolved using several SMR levels.  The number of SMR levels for each simulation is given in Table~\ref{tab:params}.  The simulations refined to a maximum SMR level of 6 are refined over a range of $\tilde{x}$ values given by $\tilde{x} \in [-0.03,0.03]$, at a resolution equivalent to a resolution of $16384 \times 32768$ over the entire mesh. The simulations refined to a maximum SMR level of 7 are refined over a range of $\tilde{x}$ values given by $\tilde{x} \in [-0.015,0.015]$, at a resolution equivalent to a resolution of $32768 \times 65536$ over the entire mesh.

Boundary conditions in $\hat{y}$ are free outflow. Boundary conditions in $\hat{x}$ have the magnetic field fixed to the value determined by Equation~\eqref{eq:vecpot}, the density of both fluids fixed to their initial values at the edge of the sheet, and the momentum of both fluids  set such that its values in the ghost zones are equal to its values in the last cell of the active mesh.

\begin{figure*}
\centering
\includegraphics[width=8cm]{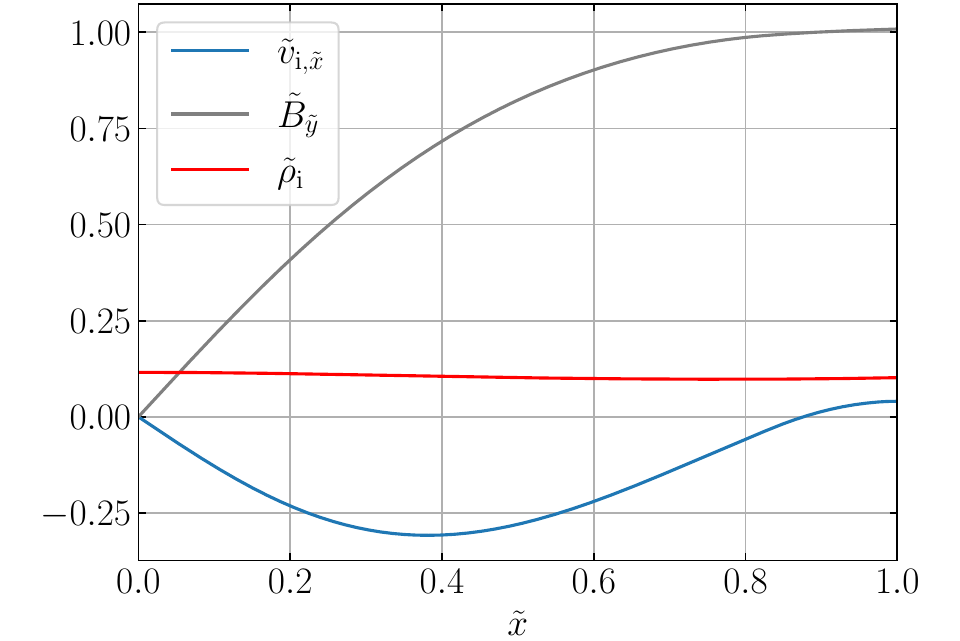}
\includegraphics[width=8cm]{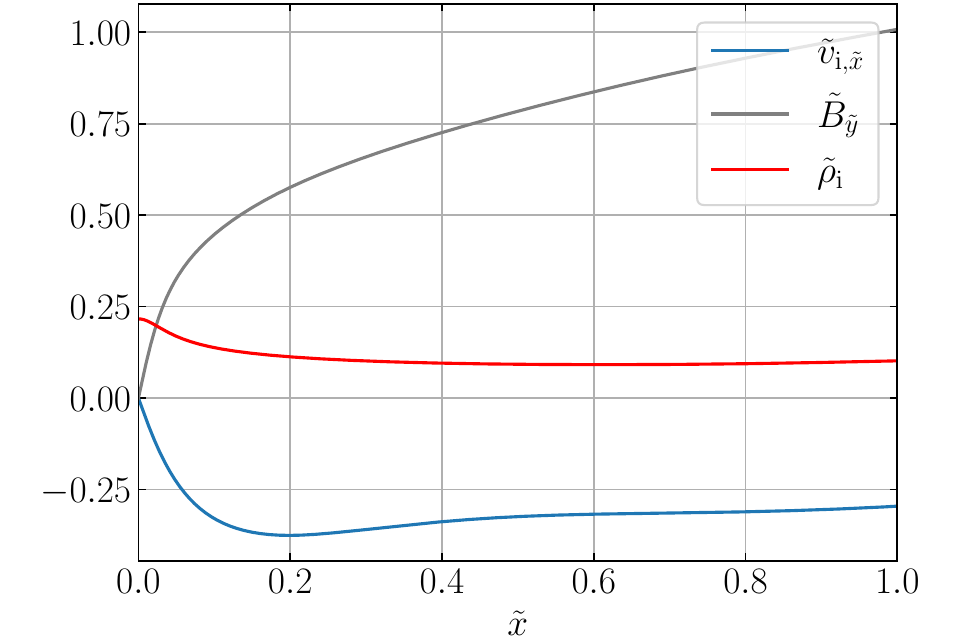}
  \caption{Early evolution of magnetic field, $x$-directed ionized fluid velocity, and ionized fluid density in the $\tilde{\alpha} = 10$ simulation, demonstrating CS formation via ambipolar diffusion. Profiles are measured at $\tilde{y}=0$ and $\tilde{t}=0.05$ (left panel) or $\tilde{t}=0.25$ (right panel).}
\label{fig:formsim}
\end{figure*}

\begin{figure*}
\centering
\subfigure{{\includegraphics[width=5.8cm]{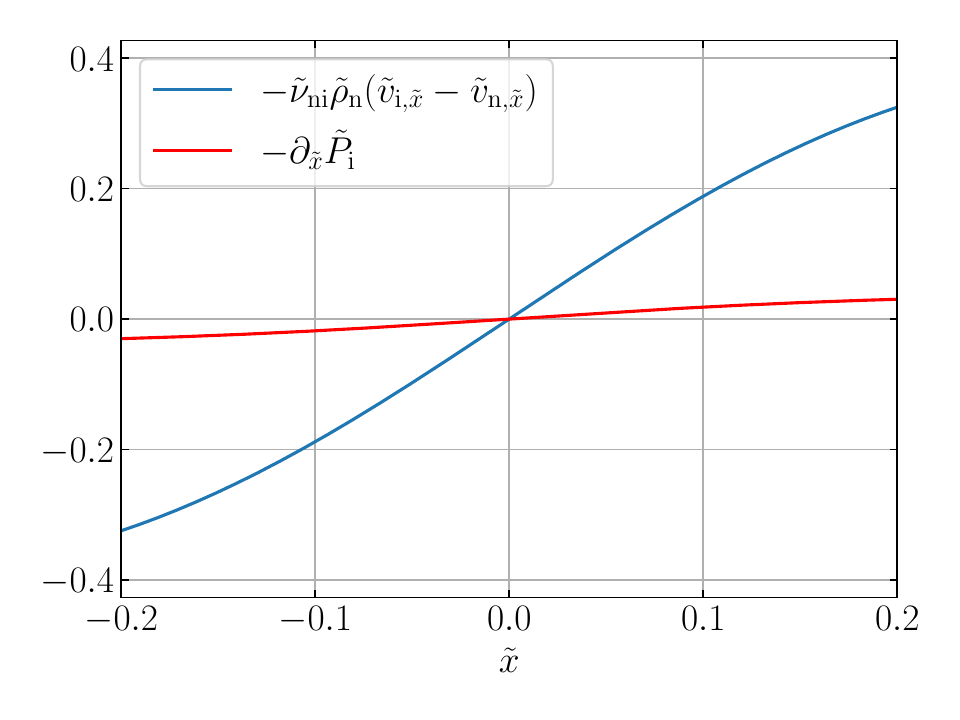} }}
\subfigure{{\includegraphics[width=5.8cm]{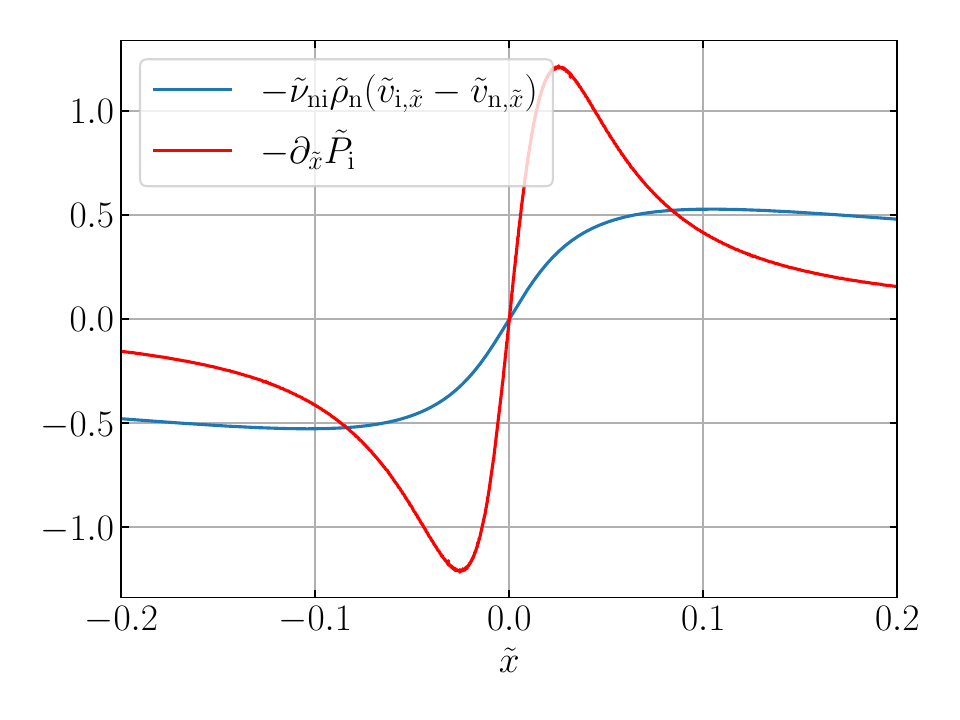} }}
\subfigure{{\includegraphics[width=5.8cm]{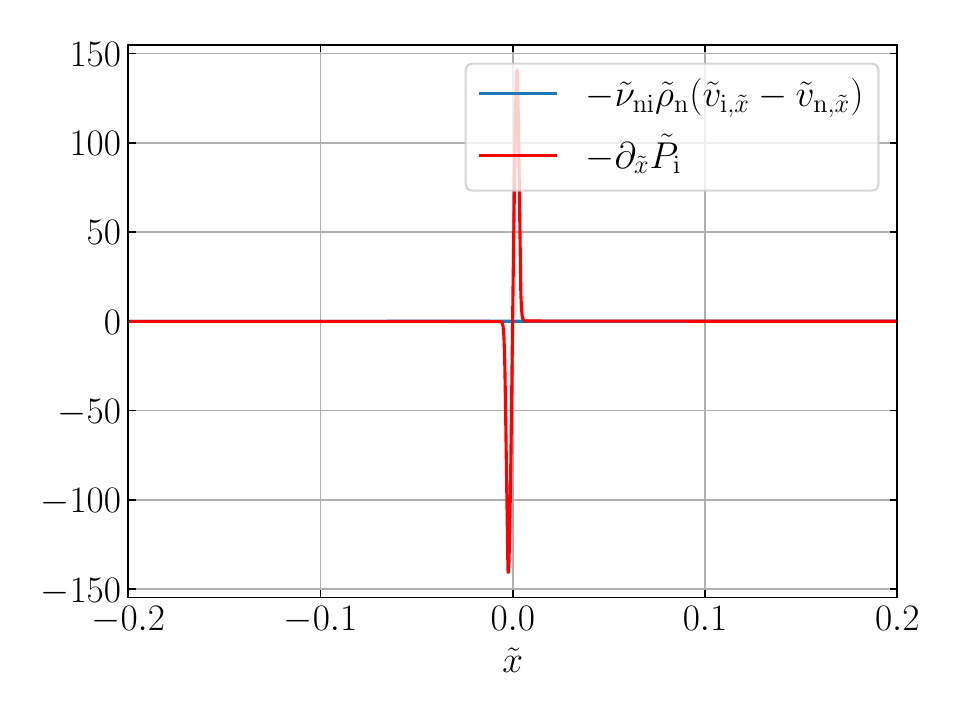} }}
  \caption{Neutral--ion drag and ionized-fluid-pressure-gradient forces (see Equation~\eqref{eq:momi}) for $\tilde{\alpha} =10$ at (left) $\tilde{t} = 0.05$, (center) $\tilde{t} = 0.25$, and (right) $\tilde{t} = 1.6$, illustrating the increasing role of the ionized fluid pressure gradient as the strong coupling breaks.}
\label{fig:breaking}
\end{figure*}

The range of parameters that can be effectively simulated is limited.  At high ionization fractions (equivalent for fixed ionization coefficient $\tilde{\xi}$ to low values of recombination coefficient $\tilde{\alpha}$), the strong-coupling assumption does not hold. At low ionization fractions (high values of $\tilde{\alpha}$), the scale of the reconnecting CS is very small, making resolved simulations prohibitively expensive. Thus, we complete a scan over a limited set of ionization and recombination rates, namely, $\tilde{\xi}=0.1$ and $\tilde{\alpha} = 10$, 16, 22, 28, 34, 40, and 60; the resulting ionization fractions at the edge of the CS are  given in Table~\ref{tab:params}.

\subsection{Evolution of Simulations}

At the start of all our simulations, the magnetic-field profile begins to sharpen via ambipolar diffusion at an increasing rate. This evolution is highlighted in Figure~\ref{fig:formsim} for the simulation with $\tilde{\alpha}  = 10$. As this process occurs, the ionized fluid pressure gradient at the center of the CS increases in magnitude, as shown in Figure~\ref{fig:breaking}.

\begin{figure}
\centering
  \includegraphics[width=\columnwidth]{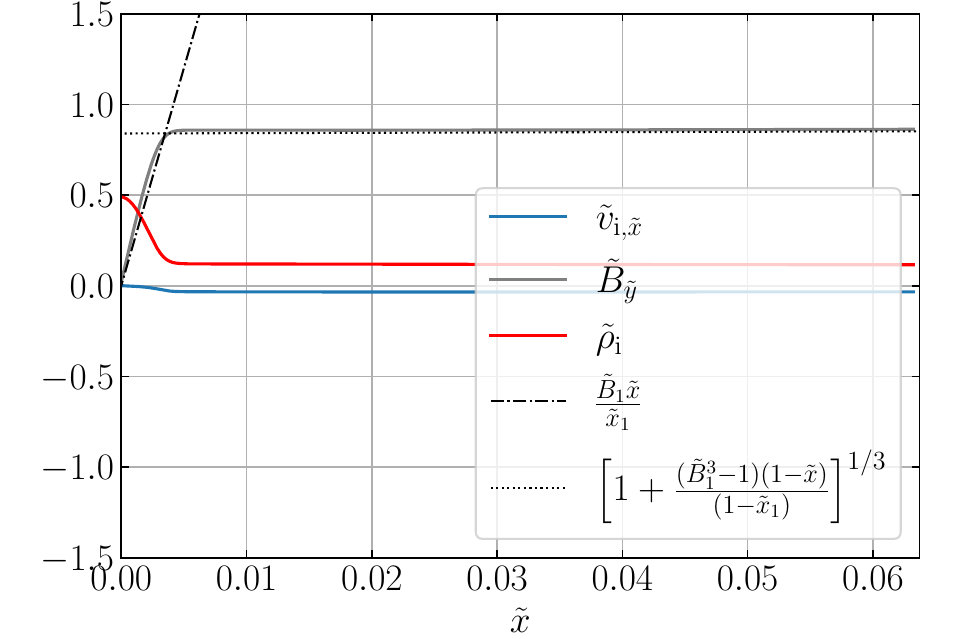}
  \caption{CS structure in the $\tilde{\alpha}=10$ simulation at $\tilde{t} = 1.6$, after the strong-coupling approximation breaks down near the center of the sheet.}
\label{fig:prof}
\end{figure}

Eventually, the increasing ionized fluid pressure gradient near the center of the CS breaks the strong-coupling assumption and a CS of the kind described in Section~\ref{sec:pisp} forms (Figure~\ref{fig:prof}): an outer region where the strong-coupling approximation holds meets an inner region where it is broken and the profile steepens to become linear. In the $\tilde{\alpha} = 10$ simulation, the boundary between these two regions is measured to occur at $\tilde{x}_1= 0.0035$, where the reconnecting magnetic field $\tilde{B}_1=0.84$.

\begin{figure}
\centering
\includegraphics[width=8cm]{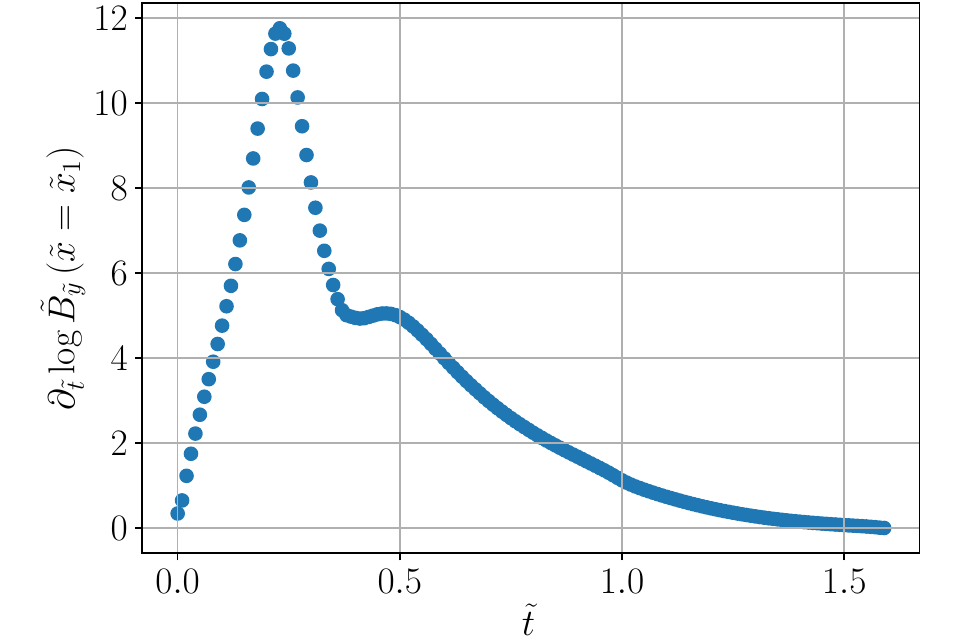} 
  \caption{Rate of change of the magnetic field at the location $\tilde{x} = \tilde{x}_1$ vs.~time, illustrating how the strong-coupling approximation breaks down in the $\tilde{\alpha} = 10$ simulation.}
\label{fig:breaking2}
\end{figure}

Figure~\ref{fig:breaking2} demonstrates how the magnetic profile changes in time during the CS formation process by displaying the rate of change of the reconnecting field, $\partial_{\tilde{t}} \log\tilde{B}_{\tilde{y}}$, evaluated at $\tilde{x}=\tilde{x}_1$. 
Initially, this quantity increases as the width of the CS decreases nonlinearly due to ambipolar diffusion. However, when the ionized fluid pressure gradient starts to become important (in the $\tilde{\alpha}$ simulation, around $\tilde{t} = 0.25$), the rate of change of $\tilde{B}_{\tilde{y}}$ stops increasing and subsequently decreases towards zero.

\begin{figure}
\centering
\subfigure{{\includegraphics[width=0.9\columnwidth]{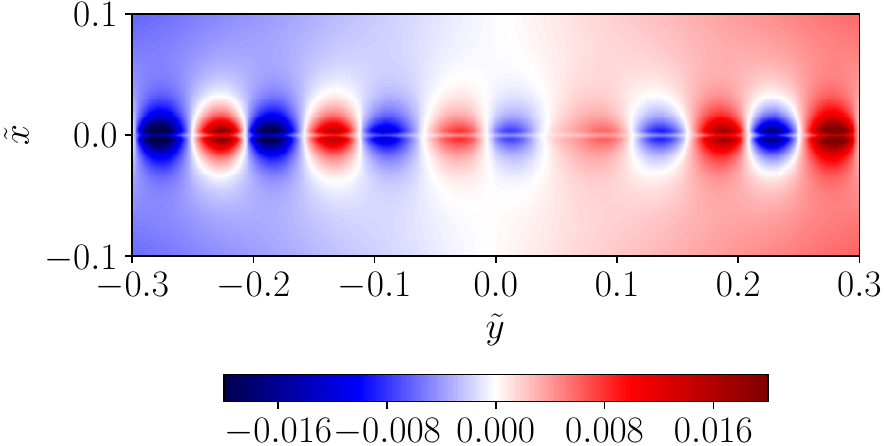} }}
\subfigure{{\includegraphics[width=0.9\columnwidth]{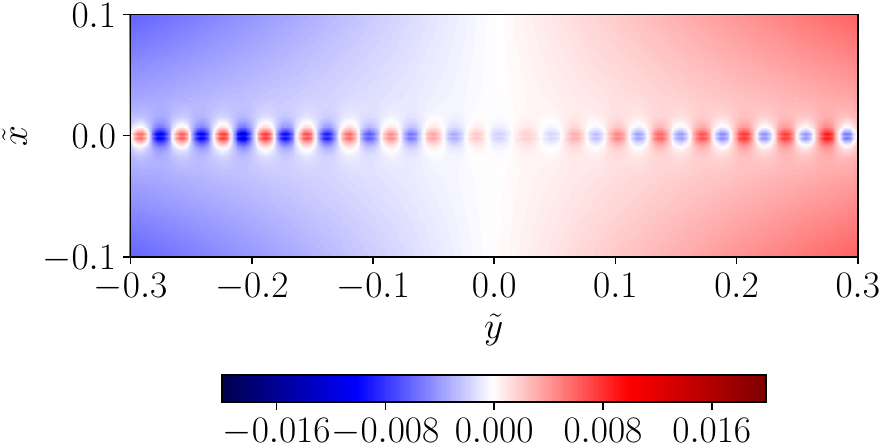} }}
  \caption{Pseudo-color plots of  $\tilde{B}_{\tilde{x}}$ from the simulations with (top) $\tilde{\alpha} = 10$ and (bottom) $\tilde{\alpha} = 60$, with the former taken at the same time shown in Figure~\ref{fig:prof}.}  
\label{fig:mode10}
\end{figure}

\begin{figure}
\centering
\includegraphics[width=\columnwidth]{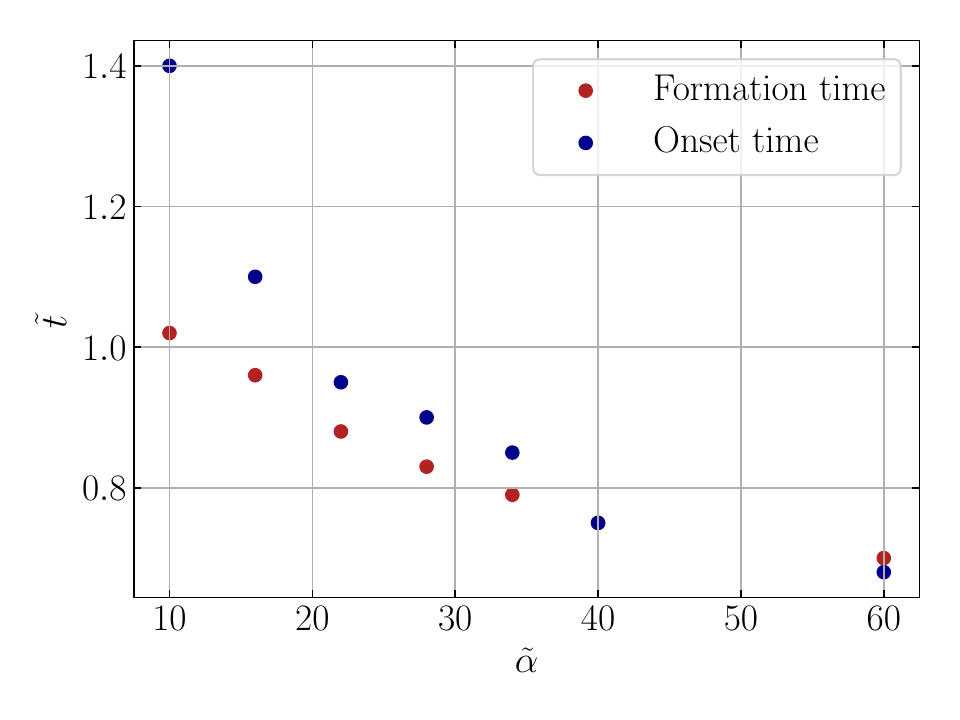}
\caption{CS-formation and tearing-onset times measured in simulations with different recombination coefficients $\tilde{\alpha}$.}
\label{fig:formtime}
\end{figure}

When the rate of change is slow enough, the CS goes unstable to the tearing instability, with tearing modes clearly visible in the profile of $\tilde{B}_{\tilde{x}}$ (Figure~\ref{fig:mode10}). 
For each of our simulations, we can estimate the tearing onset time by noting when the tearing eigenfunction is visible in $\tilde{B}_{\tilde{x}}$ evaluated at $\tilde{y} = 0$, and the CS formation time by the time at which the value of $\partial_{\tilde{t}}\log\tilde{B}_{\tilde{y}}$ decreases below unity.  These values are shown in Figure~\ref{fig:formtime}, demonstrating that the predicted condition for onset, Equation~\eqref{eq:tonset}, approximately holds.

The onset time and formation time both decrease slightly with $\tilde{\alpha}$. The decrease in the formation time is likely due to the physics of the transient phase between the second and third stages of CS formation, when the strong-coupling assumption starts to break and the final stationary CS equilibrium is approached. The decrease in the onset time is likely due to both the decrease in the formation time and the increase in the linear growth rate that occurs with $\tilde{\alpha}$. A higher linear growth rate means that tearing growth can overwhelm the CS formation earlier during the transient phase between the second and third stages. 

After tearing onsets, the simulations evolve to produce a nonlinear stochastic plasmoid chain, whose structure and evolution will be discussed in \citet{tolman23II}.

\begin{figure}
\centering
  \includegraphics[width=\columnwidth]{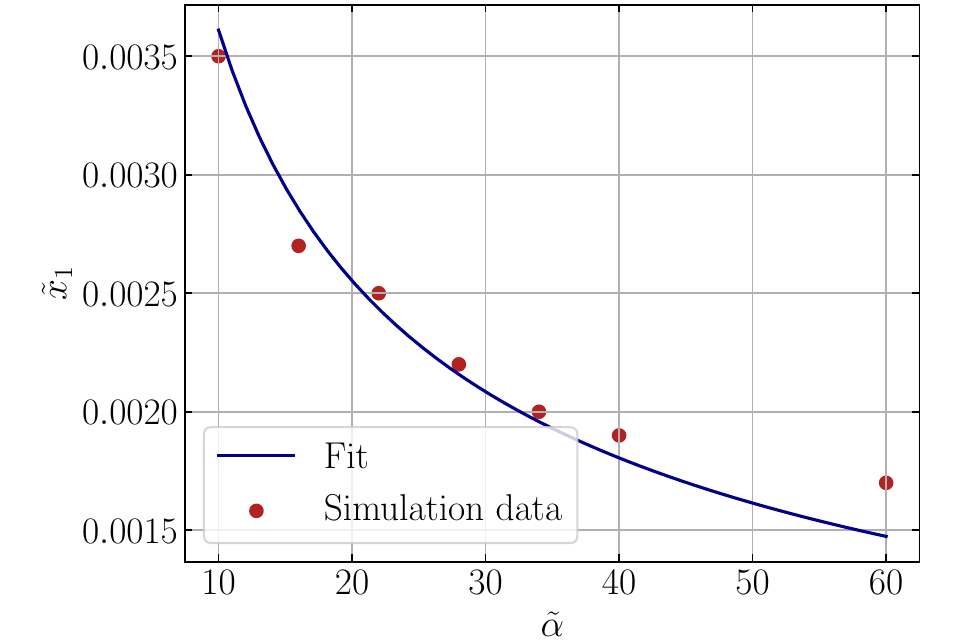}
  \caption{Values of the CS width when strong coupling breaks, $\tilde{x}_1$, measured in the simulations (red) and their least-squares fit using the predicted scaling law, Equation~\eqref{eq:x131}.}
\label{fig:x131}
\end{figure}

We can compare the wavenumber, growth rate, and eigenfunction of the linear modes observed in our simulations to the predictions obtained via the methods we have laid out in our theory. This process starts by using our simulations to verify the predicted scalings for $x_1$, $\rho_{\rm i,CS}$, and $B_1$ given in Section~\ref{sec:pisp} and to obtain the corresponding numerical proportionality coefficients. For example, we fit the scaling for $x_1$, Equation~\eqref{eq:x1}, to the values of the CS width at the time when a tearing eigenfunction is first visible in the profile of $\tilde{B}_{\tilde{x}}$ at $\hat{y} = 0$ (the value of $x_1$ is found by identifying the point where the nearly flat magnetic-field profile associated with the strong-coupling approximation intersects the inner region where the profile is linear).  Figure~\ref{fig:x131} shows that
\begin{equation}
\label{eq:x131}
    \tilde{x}_{1}   = 1.1 \, \tilde{C}_{\rm i} \sqrt{\frac{ \tilde{\eta}}{\tilde{\alpha}}}
\end{equation}
is a good fit to the simulation data.  
Likewise, we find that
\begin{equation}
\label{eq:rho31}
    \tilde{\rho}_{\rm i,CS}= 0.39 \tilde{C}_{\rm i}^{-2}
\end{equation}
and 
\begin{equation}
\label{eq:B131}
    \tilde{B}_{1} = 0.77 ,
\end{equation}
in agreement with our predictions.

\begin{figure}
\centering
  \includegraphics[width=\columnwidth]{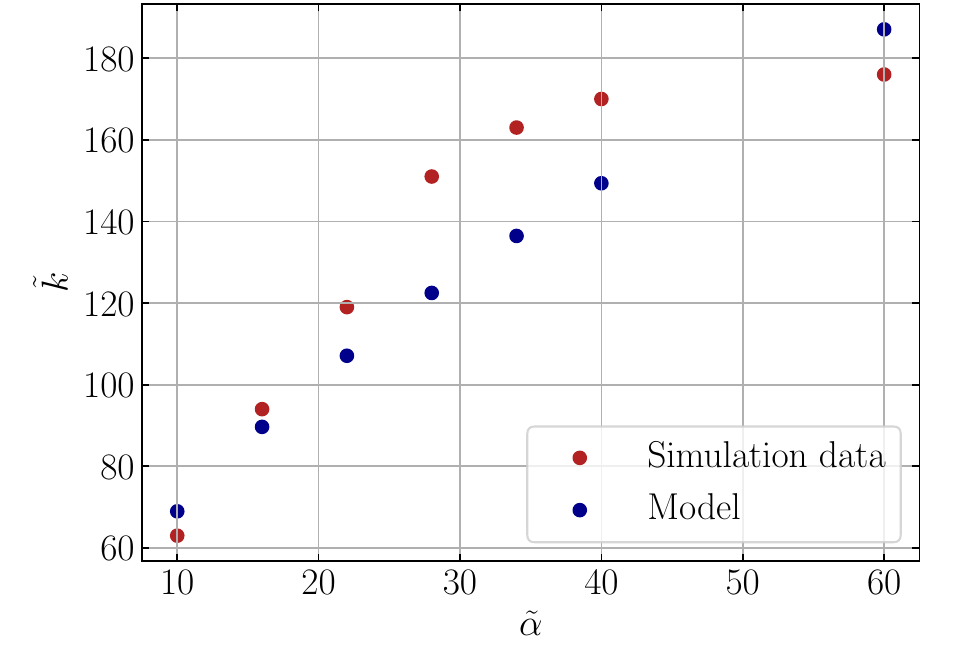}
  \caption{Predicted wavenumbers of the onset tearing mode (blue) compared to the dominant wavenumbers measured in the simulations (red), for each $\tilde{\alpha}$ listed in Table~\ref{tab:params}.}
\label{fig:31comp}
\end{figure}

Using these expressions, we can make a comparison of the most unstable mode predicted via the spectra described in Section~\ref{sec:initonset} to those observed in the simulation. This comparison is shown in Figure~\ref{fig:31comp}.  The comparison seems quite favorable given that the higher-$\tilde{\alpha}$ simulation may not be fully resolved, that the higher-$\tilde{\alpha}$ simulation approaches parameters where Equation~\eqref{eq:validcond} fails and other scalings for CS parameters should be used, and that our methods for have been approximate.
\begin{figure}
\centering
  \includegraphics[width=\columnwidth]{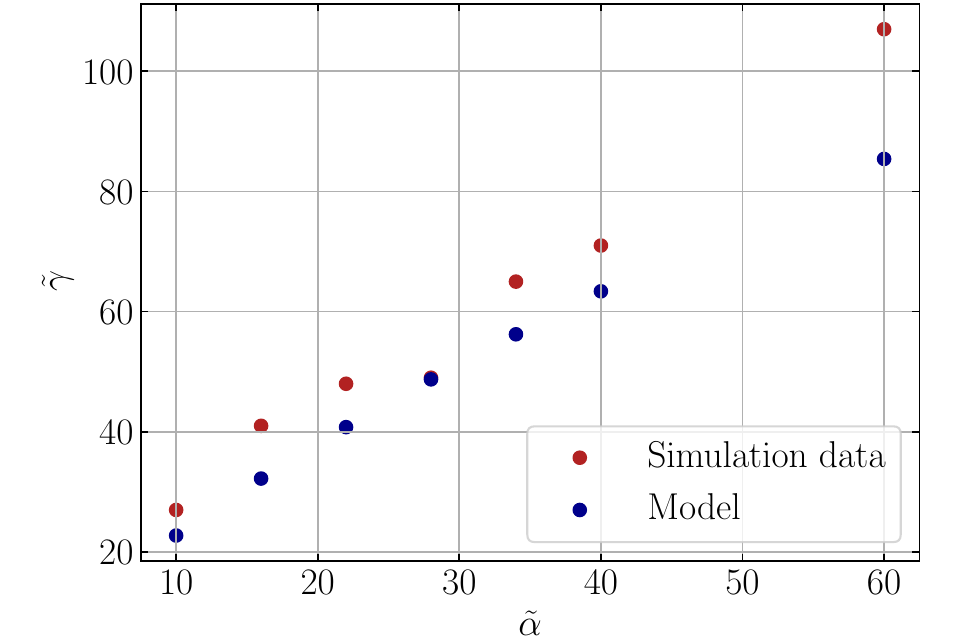}
  \caption{Predicted growth rates of the onset tearing mode (blue) compared to the growth rates measured in the simulations (red), for each $\tilde{\alpha}$ listed in Table~\ref{tab:params}.}
\label{fig:gr}
\end{figure}
We can also compare growth rates measured in the simulations (found by fitting an exponential function in time to the linear growth of the peak value of the eigenfunction of tearing, seen in $\tilde{B}_{\tilde{x}}$ evaluated at $\tilde{y} = 0$), with the analytic predictions described in Section~\ref{sec:initonset}. Figure~\ref{fig:gr} demonstrates favorable results in this regard. 

The results for mode wavenumber demonstrate less agreement than those for growth rate. This may reflect that the spectra, shown in Figure~\ref{fig:grlin}, are somewhat flat near the maximum growth rate, such that any mode near the mode with the maximum growth rate might onset.

\begin{figure}
\centering
  \includegraphics[width=\columnwidth]{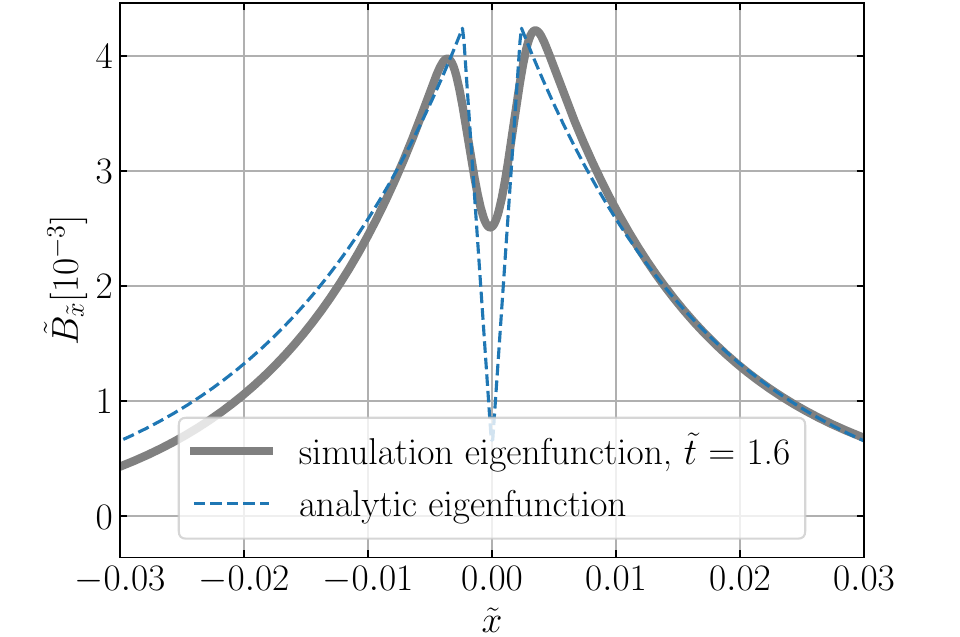}
  \caption{Profile of $\tilde{B}_{\tilde{x}}$ measured in the $\tilde{\alpha}=10$ simulation at $\tilde{t}=1.6$ compared to the analytic eigenfunction derived in Appendix~\ref{sec:deltaprime} (see Equation~\eqref{eq:psitot}).}
\label{fig:eigen}
\end{figure}

Finally, the eigenfunctions of the modes observed in the simulations also appear similar to those predicted by the theory (derived in Appendix~\ref{sec:deltaprime}). An example eigenfunction, taken from the $\tilde{\alpha} =10$ simulation at $\tilde{t}=1.6$, is provided in Figure~\ref{fig:eigen} alongside the analytically predicted eigenfunction.

\section{Conclusion}
\label{sec:conc}
The recent plasma physics literature provides an extensive discussion on the tearing-induced onset of magnetic reconnection in fully ionized plasmas. In this paper, we have presented the analogous discussion for poorly ionized plasmas, developed using analytical arguments and verified by nonlinear two-fluid MHD simulations. Our analysis yields the following important conclusions. Once the width of the forming CS shrinks below the scale $a_0$ (Equation~\eqref{eq:a0}), ambipolar diffusion causes its profile to steepen nonlinearly at an increasing rate. The consequent current singularity breaks the strong-coupling approximation within a distance $x_1$ from the null (Equation~\eqref{eq:x131}) and the CS formation rate stalls at a time $t\sim \nu^{-1}_{\rm ni0}$. Only then is the tearing instability able to onset, grow exponentially, and eventually disrupt the CS. As the plasma recombination rate increases and ionization fraction decreases, the growth rate becomes an increasing multiple of $\nu_{\rm ni0}$ and the wavelength becomes a decreasing fraction of $v_{\rm A,n0}/\nu_{\rm ni0}$. 

Our study has neglected several effects likely to be important in poorly ionized plasmas. Specifically, CSs in many poorly ionized plasmas are likely to have a guide field (that is, they will have a null in one component of the magnetic field only). The presence of such a guide field will inhibit CS formation through ambipolar diffusion \citep{zweibel1997current}, perhaps preventing the triggering of the tearing instability and certainly modifying the nature of its onset.  In addition, CS formation occurring in a system where significant turbulent motion exists might occur in a different manner than the one we have described. Tearing onset in these situations is a fruitful avenue for future work. 

In addition, poorly ionized plasmas can be subject to the Hall effect, which becomes important on scales larger than in a fully ionized plasma because the ions are weighed down relative to the electrons not only by their own inertia but also by the inertia of the neutral species to which they are collisionally coupled \citep[e.g.,][]{pandey2008hall}. This effect may be important in the systems that we study and deserves further investigation.  Finally, we have taken an isothermal equation of state, neglecting any heating that may occur during the reconnection process (both ambipolar and Ohmic). We expect that this final assumption will have limited effect on the linear stage, though it may become important in the nonlinear stage, e.g., if the heating becomes large enough to thermally ionize the plasma and thereby change the degree of ionization appreciably. This will be checked in subsequent work \citep{tolman23II}.

Nonetheless, we can identify areas where our work is likely to advance our understanding of astrophysical plasmas. For example, one frontier area in the research of fully ionized plasmas concerns the impact of tearing modes on MHD turbulence. Arguments similar in spirit to those employed here have been used to predict a range of wavenumbers where the turbulent cascade of energy from large to small scales is mediated by reconnection \citep{loureiro2017role}.  We anticipate that similar physics could be at play in poorly ionized turbulent plasmas, the most immediate applications being to protoplanetary disks and the colder phases of the interstellar medium. Concerning the latter, the structures produced by tearing onset could affect the transport of cosmic rays and the scintillation of pulsar signals \citep{stinebring2022scintillation,fielding2022plasmoid,hopkins2022standard,kempski2022reconciling,reardon2023determining}. Finally, our work complements multi-fluid numerical studies of magnetic reconnection in the low solar atmosphere \citep[e.g.,][]{leake2013magnetic,Ni_2018}, in which the ionization fraction varies from ${\sim}10^{-4}$ in the photosphere to ${\sim}1$ at the top of the chromosphere and explosive events such as jets and flares occur that have been attributed to magnetic reconnection. With our analysis of the onset and linear evolution of tearing-mediated reconnection in poorly ionized CSs complete, understanding its nonlinear evolution and the impact on turbulent cascades and chromospheric dynamics is the clear next step.

\section*{Acknowledgements}
We are grateful for helpful conversations with Amitava Bhattacharjee, Alexander Chernoglazov, Yuri Levin, Nuno Loureiro, Brian Metzger, Alexander Philippov, Bart Ripperda, George Wong, Muni Zhou, and Ellen Zweibel. ET was supported by the W.M.~Keck
Foundation Fund at the Institute for Advanced Study. Research at the Flatiron Institute is supported by the Simons Foundation.

\appendix 

\section{Adaptation of fully ionized tearing growth rate to a poorly ionized plasma}
\label{sec:stationary}

Section~\ref{sec:tearing} provided results from the linear theory of the tearing mode in a stationary, poorly ionized plasma. These results were then used in Section~\ref{sec:onset} to predict the time of tearing onset and the dominant destabilizing tearing mode. Here we detail the derivation of this linear theory. 

We begin by adopting a background state that is composed of a stationary and uniform poorly ionized plasma and threaded by a magnetic field satisfying $\bb{B}_0 = B_y(x)\ey$ with $B_y(0)=0$. To this background we apply small-amplitude perturbations ($\delta\bb{B}$, $\delta\bb{v}_{\rm i}$, $\delta\bb{v}_{\rm n}$, etc.)~and take their time dependence to be ${\propto}\exp(\gamma t)$, where $\gamma$ is the growth rate. Substituting this decomposition into the momentum equations for the neutral and ionized fluids, Equations~\eqref{eq:momn} and \eqref{eq:momi} respectively, and retaining only those terms that are first order in the perturbation amplitudes yields the following linearized equations:

\begin{equation}
    \gamma \rho_{\rm n}\delta\bb{v}_{\rm n} = -\grad \delta P_{\rm n} - \nu_{\rm ni}\rho_{\rm n}(\delta\bb{v}_{\rm n}-\delta\bb{v}_{\rm i}) - \xi\rho_{\rm n}\delta\bb{v}_{\rm n} + \alpha\rho^2_{\rm i}\delta\bb{v}_{\rm i} , 
\end{equation}
\begin{equation}
    \gamma \rho_{\rm i}\delta\bb{v}_{\rm i} = -\grad \delta P_{\rm i} + (\grad\btimes\delta\bb{B})\btimes\bb{B}_0 + (\grad\btimes \bb{B}_0)\btimes\delta\bb{B} - \nu_{\rm ni}\rho_{\rm n}(\delta\bb{v}_{\rm i}-\delta\bb{v}_{\rm n}) + \xi\rho_{\rm n}\delta\bb{v}_{\rm n} - \alpha\rho^2_{\rm i}\bb{v}_{\rm i}. 
\end{equation}
The thermal pressures may be eliminated by taking the curl of both equations, which may then be rearranged to become
\begin{equation}
    (\gamma\rho_{\rm n}+\nu_{\rm ni}\rho_{\rm n}+\xi\rho_{\rm n}) \grad\btimes\delta\bb{v}_{\rm n} = (\nu_{\rm ni}\rho_{\rm n}+\alpha\rho^2_{\rm i})\grad\btimes\delta\bb{v}_{\rm i},
\end{equation}
\begin{equation}
    (\gamma\rho_{\rm i}+\nu_{\rm ni}\rho_{\rm n}+\alpha\rho^2_{\rm i})\grad\btimes\delta\bb{v}_{\rm i} = (\nu_{\rm ni}\rho_{\rm n}+\xi\rho_{\rm n})\grad\btimes\delta\bb{v}_{\rm n} + \grad\btimes[(\bb{B}_0\bcdot\grad)\delta\bb{B} + (\delta\bb{B}\bcdot\grad)\bb{B}_0] .
\end{equation}
These two equations may then be combined by eliminating $(\grad\btimes\delta\bb{v}_{\rm n})$ to obtain

\begin{equation}\label{eq:momeq}
    (\grad\times\delta\bb{v}_{\rm i} ) \,\gamma \rho_{\rm i} 
        \left(1 + \frac{\rho_{\rm n}}{\rho_{\rm i}} \frac{\nu_{\rm ni} + \alpha \rho_{\rm i}^2/\rho_{\rm n}}{\gamma +\nu_{\rm ni} +\xi } \right)
    = \grad\btimes[(\bb{B}_0\bcdot\grad)\delta\bb{B} + (\delta\bb{B}\bcdot\grad)\bb{B}_0] .
\end{equation}
Defining an effective density
\begin{equation}
    \rho_{\rm i}' \equiv \rho_{\rm i}\left(1 + \frac{\rho_{\rm n}}{\rho_{\rm i}} \frac{\nu_{\rm ni} + \alpha \rho_{\rm i}^2/\rho_{\rm n}}{\gamma +\nu_{\rm ni} +\xi } \right),
\end{equation}
we see that Equation~\eqref{eq:momeq} may be written compactly as
\begin{equation}
    (\grad\btimes\delta\bb{v}_{\rm i}) \, \gamma\rho'_{\rm i} =   \grad\btimes[(\bb{B}_0\bcdot\grad)\delta\bb{B} + (\delta\bb{B}\bcdot\grad)\bb{B}_0] .
\end{equation}
This equation is equivalent to its fully ionized counterpart but for the modified ionized density, demonstrating that the effect of the neutrals on the tearing instability is to weigh down motions in the ionized species through collisions and recombination. Following \citet{zweibel1989magnetic}, who investigated linear tearing in a poorly ionized plasma without including ionization and recombination, we introduce the effective Alfv\'en velocity
\begin{equation}
    \bb{v}'_{\rm A,i} \equiv \frac{\bb{B}_0}{\sqrt{\rho'_{\rm i}}} = v_{\rm A,i}  \left(1 + \frac{\rho_{\rm n}}{\rho_{\rm i}} \frac{\nu_{\rm ni} + \alpha \rho_{\rm i}^2/\rho_{\rm n}}{\gamma +\nu_{\rm ni} +\xi } \right)^{-1/2},
\end{equation}
and simply replace the Alfv\'en-crossing time of the perturbation's wavelength that appears in the fully ionized dispersion relation~\eqref{eq:gamma} as follows: $\tau_{\rm A}  \rightarrow \tau'_{\rm A,i} \equiv (kv'_{\rm A,i})^{-1}$. The  linear tearing theory then proceeds in the usual way \citep[e.g.,][]{furth1963finite,coppi1966resistive,boldyrev2018calculations}.

\section{Derivation of the Tearing Stability Index}
\label{sec:deltaprime}

To obtain a quantitative value for the tearing growth rate, an expression for the tearing-mode stability index $\Delta'$ in Equation~\eqref{eq:gamma} is required. This index depends on the functional form of the CS profile. In studies of fully ionized plasmas, this profile is often taken to be the \citet{harris1962plasma} profile, for which $\Delta' a = 2(1/ka - ka)$. Here we derive an expression for the stability index of the poorly ionized Sweet--Parker profile, Equation~\eqref{eq:btot}. 

We begin by calculating the tearing-mode flux eigenfunction $\Psi(x)$ in Equation~\eqref{eq:eigdef} by solving the outer-region tearing equation (for a pedagogical introduction to this equation, see \cite{boldyrev2018calculations}),
\begin{equation}
\label{eq:diffeq}
    \DD{x}{\Psi} = \left( k^2 + \frac{B''_y}{B_y} \right) \Psi ,
\end{equation}
where $B_y(x)$ is given by Equation~\eqref{eq:btot} and the prime denotes differentiation with respect to $x$; the result is then substituted into the definition
\begin{equation}
\label{eq:deltapr}
   \Delta' \equiv \frac{1}{\Psi(0)} \left[\D{x}{\Psi(0^+)} - \D{x}{\Psi(0^-)}\right].
\end{equation}
For this $B_y(x)$, the value of $|B''_y/B_y|$ is sharply peaked at the characteristic width of the CS, $x = \pm x_1$. Therefore, we make the approximation
\begin{equation}
\label{eq:deltaexpr}
    \frac{B''_y}{B_y} \approx - \chi\, \delta(x-x_1) - \chi\, \delta(x+x_1) .
\end{equation}
The parameter $\chi$ is determined by integrating Equation~\eqref{eq:deltaexpr} about $x=x_1$ to find
\begin{equation}
    B'_y(x^+_1) - B'_y(x^-_1)= -\chi B_y(x_1).
\end{equation}
Then, use of Equation~\eqref{eq:btot} for the CS profile gives 
\begin{equation}
\label{eq:alphadef}
    \chi = \frac{1}{x_1} - \frac{(B_0/B_1)^3-1}{3(a_0-x_1)} \approx \frac{1}{x_1} ,
\end{equation}
where the approximation, valid for $x_1 \ll a_0$, is taken to prevent unreasonably complicated expressions later on. We must then solve Equation~\eqref{eq:diffeq} with Equation~\eqref{eq:deltaexpr} having $\chi\approx x^{-1}_1$. 
The general solution must go to zero as $|x|\rightarrow\infty$, and so
\begin{equation}
\label{eq:psitot}
   \Psi( x) \propto
        \left\{ \begin{array}{ll}
            \rme^{kx}, &x<-x_1 \\
            C_1 \rme^{-kx} + C_2 \rme^{kx}, &-x_1<x<0 \\
            C_3 \rme^{-kx} + C_4 \rme^{kx}, &0<x< x_1 \\
            C_5 \rme^{-k x} , & x>x_1
        \end{array} \right..
\end{equation}
We have left the overall normalization free.  To find the constants in Equation~\eqref{eq:psitot}, we impose continuity of $\Psi(x)$ at $x = -x_1$, $0$, and $x_1$ and use Equations~\eqref{eq:diffeq} and~\eqref{eq:deltaexpr} to constrain the discontinuity in $\Psi'(x)$ at $x=\pm x_1$. 
Applying these conditions and using Equation~\eqref{eq:alphadef} we find the following:
\begin{equation}
    C_1 = C_4 = \frac{\rme^{-2kx_1}}{2kx_1} , \quad C_2 = C_3 = 1 - \frac{1}{2kx_1}, \quad C_5 = 1 . 
\end{equation}
With these coefficients determined, we can evaluate Equation~\eqref{eq:deltapr} to obtain
\begin{equation}
\label{eq:dpvalapp}
    \Delta'(k) = k\left[-2 + \frac{4}{1 +e^{2 k x_1}\left(2 k x_1 -1 \right)} \right] .
\end{equation}
For small $kx_1$, this approaches $\Delta' \approx (kx_1^2)^{-1}$, which agrees with a trend found numerically by \cite{brandenburg1995effects}. Note that this long-wavelength limit is the same as that corresponding to a \citet{harris1962plasma} equilibrium having width $x_1$.

\section{Typical Values for Key Parameters}
\label{sec:typval}
In this Appendix we provide typical values for various key parameters that appear in our calculations. These values are based on fiducial conditions in star-forming molecular clouds, an example poorly ionized system in which magnetic reconnection might be important (though they may be readily scaled for application to other systems). Molecular clouds contain gas that is cold ($T\approx 10~{\rm K}$) and composed mostly of neutral molecular hydrogen H$_2$ with 20\% He by number, trace amounts of electrons and molecular and atomic ions (e.g., HCO$^+$, Na$^+$, Mg$^+$), and ${\sim}1\%$ by mass of dust grains (both charged and neutral). The neutral fluid mass density is
\begin{equation}
    \rho_{\rm n} = 1.2\times 10^{-20} \, \biggl(\frac{n_{\rm n}}{3\times 10^3~{\rm cm}^{-3}}\biggr)~{\rm g~cm}^{-3} ,
\end{equation}
where $m_{\rm p}$ is the proton mass and $m_{\rm n}=2.33m_{\rm p}$ is the mean mass per neutral particle. A typical Alfv\'{e}n velocity in the neutrals is then
\begin{equation}\label{eq:vAnapp}
    v_{\rm A,n} = 4\times 10^4 \, \biggl(\frac{B}{15~\mu{\rm G}}\biggr) \biggl(\frac{n_{\rm n}}{3\times 10^3~{\rm cm}^{-3}}\biggr)^{-1/2} ~{\rm cm~s}^{-1},
\end{equation}
comparable to the typical turbulent velocities inferred from line broadening. Given fiducial values for the cosmic-ray ionization rate $\zeta_{\rm cr}$ and dissociative recombination rate $\alpha_{\rm DR}$ \citep[e.g.,][]{umebayashi1990magnetic}, the inverse-mass-weighted ionization and recombination coefficients that feature in Equations~\eqref{eq:contn} and \eqref{eq:conti} are
\begin{gather}
    \xi = \zeta_{\rm CR} \frac{m_{\rm i} n_{\rm H_2}}{\rho_{\rm n}} = 5.2\times 10^{-16} \, \biggl(\frac{\zeta_{\rm CR}}{5\times 10^{-17}~{\rm s}^{-1}}\biggr) \biggl(\frac{m_{\rm i}}{29m_{\rm p}}\biggr) ~{\rm s}^{-1} , \\
    \alpha = \alpha_{\rm DR} \frac{1}{m_{\rm i}} = 5.2\times 10^{16} \, \biggl(\frac{\alpha_{\rm DR}}{2.5\times 10^{-6}~{\rm cm}^3~{\rm s}^{-1}}\biggr) \biggl(\frac{T}{10~{\rm K}}\biggr)^{-3/4} \biggl(\frac{m_{\rm i}}{29m_{\rm p}}\biggr)^{-1}~{\rm cm}^3~{\rm g}^{-1}~{\rm s}^{-1} .
\end{gather}
Here we have taken the dominant ion mass to be that of HCO$^+$ ($29m_{\rm p}$); other relevant ion species include Na$^+$ ($23m_{\rm p}$) and Mg$^+$ ($24m_{\rm p}$), for which radiative recombination rates should be used. Using these fiducial values, ionization equilibrium (Equation~\eqref{eq:ionequil}) then provides
\begin{equation}\label{eq:ionequilapp}
    \frac{\rho_{\rm i}}{\rho_{\rm n}} = 9.3\times 10^{-7} \, \biggl(\frac{n_{\rm n}}{3\times 10^3~{\rm cm}^{-3}}\biggr)^{-1/2} \biggl(\frac{T}{10~{\rm K}}\biggr)^{3/8} \biggl(\frac{m_{\rm i}}{29m_{\rm p}}\biggr).
\end{equation}
The mean collisional rate $\langle\sigma w\rangle_{\rm in}$ is equal to $1.69\times 10^{-9}~{\rm cm}^3~{\rm s}^{-1}$ for HCO$^+$--H$_2$ collisions, and almost identical to this value for Na$^+$--H$_2$ and Mg$^+$--H$_2$ collisions; the presence of He effectively reduces this value by a factor of 0.81 by lengthening the slowing-down time relative to the value it would have if only HCO$^+$--H$_2$ collisions were considered \citep{mouschovias96}. Substituting these values into Equation~\eqref{eq:nuni}, adopting ionization equilibrium, and setting $m_{\rm i}=29m_{\rm p}$ yields 
\begin{equation}\label{eq:nuniapp}
    \nu_{\rm ni} = 2.9\times 10^{-13}\, \biggl(\frac{n_{\rm n}}{3\times 10^3~{\rm cm}^{-3}}\biggr)^{1/2} \biggl(\frac{T}{10~{\rm K}}\biggr)^{3/8}~{\rm s}^{-1},
\end{equation}
or $9.1~{\rm Myr}^{-1}$. The ambipolar-diffusion length scale (Equation~\eqref{eq:a0}) that results by taking the ratio of Equations~\eqref{eq:vAnapp} and \eqref{eq:nuniapp} is
\begin{equation}
    a_0 = 1.4\times 10^{17} \, \biggl(\frac{B}{15~\mu{\rm G}}\biggr) \biggl(\frac{n_{\rm n}}{3\times 10^3~{\rm cm}^{-3}}\biggr)^{-1} \biggl(\frac{T}{10~{\rm K}}\biggr)^{-3/8}~{\rm cm} ,
\end{equation}
or $0.044~{\rm pc}$. Finally, at molecular cloud densities and temperatures, the electrical resistivity is dominated by collisions between the electrons and the neutrals, for which $\nu_{\rm en} = \rho_{\rm n}\langle\sigma w\rangle_{\rm en}/(m_{\rm e}+m_{\rm H_2})$ with $\langle\sigma w\rangle_{\rm en}=1.1\times 10^{-9}~(T/10~{\rm K})^{1/2}~{\rm cm}^3~{\rm s}^{-1}$ \citep{mouschovias96}. Thus,
\begin{equation}
    \eta = 4.7\times 10^9 \, \biggl(\frac{n_{\rm n}}{3\times 10^3~{\rm cm}^{-3}}\biggr)^{1/2} \biggl(\frac{T}{10~{\rm K}}\biggr)^{1/8} ~{\rm cm}^2~{\rm s}^{-1} .
\end{equation}
We can now use the above physical parameters to estimate the sizes of the following dimensionless free parameters:
\begin{gather}
    \tilde{\xi} \approx 2\times 10^{-3} \, \biggl(\frac{n_{\rm n}}{3\times 10^3~{\rm cm}^{-3}}\biggr)^{-1/2}\biggl(\frac{T}{10~{\rm K}}\biggr)^{-3/8} , \\
    \tilde{\alpha} \approx 2\times 10^9 \, \biggl(\frac{n_{\rm n}}{3\times 10^3~{\rm cm}^{-3}}\biggr)^{1/2} \biggl(\frac{T}{10~{\rm K}}\biggr)^{-9/8} ,\\
    \tilde{\eta} = S_{a_0}^{-1} \approx 9\times 10^{-13} \, \biggl(\frac{B}{15~\mu{\rm G}}\biggr)^{-2} \biggl(\frac{n_{\rm n}}{3\times 10^3~{\rm cm}^{-3}}\biggr)^{2} \biggl(\frac{T}{10~{\rm K}}\biggr)^{1/2} ,\\
    \mathcal{Z} \approx 10^{-7} \biggl(\frac{T}{10~{\rm K}}\biggr)^{29/16}  \biggl(\frac{B}{15~\mu{\rm G}}\biggr)^{-2} \biggl(\frac{n_{\rm n}}{3\times 10^3~{\rm cm}^{-3}}\biggr)^{-1/4} .
\end{gather}
These values indicate that actual CSs in star-forming molecular clouds are likely to be far too small to resolve numerically.

\bibliography{bib}{}
\bibliographystyle{aasjournal}

\end{document}